# The dynamics of polymers in solution with hydrodynamic memory[*]

V. Lisy, J. Tothova, and B. Brutovsky
*Institute of Physics, P.J. Safarik University, Jesenna 5, 041 54 Kosice, Slovakia*

A.V. Zatovsky
*Department of Theoretical Physics, I.I. Mechnikov Odessa National University, 2, Dvoryanskaya Str., 65026 Odessa, Ukraine*

**ABSTRACT**

The dynamics of individual polymers in solution is fundamental for understanding the properties of polymeric systems. Consequently, considerable work has been devoted towards understanding polymer dynamics. In spite of the long-standing investigations, a number of problems remains between the theory and experiments, such as the dynamic light or neutron scattering. For example, the value of the first cumulant of the dynamic structure factor (DSF) is lower than the theoretical prediction for flexible polymers. The origin of these discrepancies is a matter of continuous discussion. The development of the theory of polymer dynamics is thus still of interest.

This work represents an attempt of such a development. The main idea of the proposed generalization comes from the theory of the Brownian motion, which lies in the basis of the bead-spring models of polymer dynamics. In the classical (Einstein) description the resistance force on the bead moving in a liquid is the Stokes force, which is valid only for the steady motion. In a more general nonstationary case the hydrodynamic memory, which is a consequence of fluid inertia, should be taken into account. In the Brownian motion it displays in the famous "long-time tails" in the velocity autocorrelation function. The time dependence of the mean square displacement (MSD) of the particle changes from the "ballistic" regime at short times to the Einstein diffusion at long times. We have found similar effects in the dynamics of polymers. We give the corresponding generalization of the Rouse-Zimm (RZ) theory. It is shown that the time correlation functions describing the polymer motion essentially differ from those in the RZ models. The MSD of the polymer coil is at short times proportional to $t^2$ (instead of $t$). At long times it contains additional (to the Einstein term) contributions, the leading of which is $\sim t^{1/2}$. The relaxation of the internal normal modes of the polymer differs from the traditional exponential decay. This is displayed in the tails of their correlation functions, the longest-lived being $\sim t^{-3/2}$ in the Rouse limit and $t^{-5/2}$ in the Zimm case when the hydrodynamic interaction is taken into account. It is discussed that the found peculiarities, in particular a slower diffusion of the coil, should be observable in dynamic scattering experiments. The DSF and the first cumulant of the polymer coil are calculated. Finally, we extend the theory to the situation when the dynamics of the studied polymer is influenced by the presence of other polymers in dilute solution.

---

**CONTENTS**

# INTRODUCTION

The richness of possible structures and dynamical behavior of polymeric systems makes them attractive subject in the physics of soft condensed matter. Understanding the properties of these systems is based on understanding the static and dynamic properties of individual polymers in solution. For example, the slow internal dynamics of long polymer chains is the origin of viscoelasticity of dilute polymer solutions [1 - 3] and the natural motions of many biological macromolecules appear to be essential for their functioning [4, 5]. Consequently, considerable experimental and theoretical work has been devoted to the study of polymer dynamics. Among the available theories the most popular are the bead-spring Rouse [6, 7] and Zimm [8] phenomenological models that are considered universal models of the long-time and scale dynamics. In spite of the long-standing investigations of these models, there are still problems in their applications in the interpretation of experiments, such as the dynamic light or neutron scattering [9 - 14]. For example, the "universal" plateau in the plot of $\Gamma/k^3$ *vs.* $kR_G$ ($\Gamma$ is the first cumulant of the dynamic structure factor (DSF), $k$ is the wave-vector transfer at the scattering, and $R_G$ the gyration radius of the polymer) is lower than predicted in the Rouse-Zimm (RZ) theory for flexible polymers. The diffusion coefficients of the polymer coils determined from the dynamic scattering at small $kR_G$ are lower than calculated from the theory using $R_G$ from the static scattering data, *etc*. The origin of these discrepancies remains unclear for many years [1, 12 - 14]. These and other problems are the matter of continuous discussion. The development of the theory of polymer dynamics, even in the simplest case of ideal flexible polymers in dilute solutions, is thus a challenge in the soft condensed matter physics.

The present work represents an attempt of such development. The generalization proposed by us can be called the hydrodynamic RZ model. The main idea comes from the theory of the Brownian motion, which lies in the basis of the RZ model. In the Einstein description (as well as in the later Ornstein-Langevin theory [15]) the resistance force on the particle during its motion in the liquid is the Stokes force, which is at a given moment of time $t$ determined by the bead velocity at the same time. This approximation is valid only for the steady motion, i.e. at $t \to \infty$ [16]. In a more general case, within the usual Navier-Stokes hydrodynamics, the friction force on the Brownian particle should be the Boussinesq force [17] called the history force since, in general, at the time $t$ it depends on the state of the particle motion in all the preceding moments of time (for incompressible fluids $t \gg b/c$ where $c$ is the velocity of sound and $b$ the radius of the spherical particle). This was noted by Vladimirsky and Terletzky [18] who have built the first hydrodynamic theory of the translational Brownian motion. For compressible fluids it was generalized in the work [19], and the hydrodynamic rotational Brownian motion was first considered by Zatovsky [20]. (The investigations [18, 20] (reviewed in Ref. [21]) remained unknown or very little known to the physical community and long time after the appearance of the original papers their main results were rediscovered by other authors.) The hydrodynamic memory (or the so-called viscous aftereffect) is a consequence of fluid inertia. In the Brownian motion it displays in the "long-time tails" in the particle velocity autocorrelation function that became famous after their discovery in the computer experiments on simple liquids (for a review see, e.g., Ref. [22]). The tails reflect strong correlations with the initial state of the particle and persist for a long time. The time dependence of the mean square displacement (MSD) of the particle changes from the "ballistic" regime at short times to the Einstein diffusion when the MSD is proportional to $t$. The nondiffusive regime with the characteristic time $\tau_b = b^2\rho/\eta$ ($\rho$ is the density of the solvent and $\eta$ its viscosity) was observed in the dynamic light scattering (DLS) experiments, e.g. [23 - 25]. For example, using the diffusive wave spectroscopy the ballistic motion of polystyrene spheres with the radius $b = 0.206$ µm in aqueous solution (with the characteristic time $\tau_b$ about 0.04 µs) was observed [25]. The size of such particles corresponds to the hydrodynamic radius of the DNA polymer coil of a molecular weight $6 \times 10^6$ g/mol. Even much smaller particles were successfully studied in other experiments and the correspondence with the theory has been demonstrated. One could thus expect that similar memory effects exist in the dynamics of polymers and are observable in the DLS (or neutron scattering) experiments. The corresponding generalization of the RZ theory was given in our recent papers [26 - 30]. This work is devoted mainly to the calculation of the DSF and its first cumulant for polymer coils in the solvent with hydrodynamic memory. It is shown how the memory effects reveal in observable quantities, such as the first cumulant. We have also derived the generalized RZ equation taking into account the effects of hydrodynamic noise, with random fluctuations of the hydrodynamic stress tensor being responsible for the noise. As a result, the spectral properties of the random forces acting on the polymer segments are not delta-correlated and are determined by the hydrodynamic susceptibility of the solvent. Using the preaveraging approximation we relate the time correlation functions of the Fourier components of the segment radius vector to the correlation functions of the hydrodynamic field created by the noise. The velocity correlation function of the center of mass of the coil has been considered in detail. At long times its behavior follows the algebraic $t^{-3/2}$ law and does not depend on the polymer parameters. Finally, we sketch the phenomenological theory that takes into account the presence of other polymers in dilute solution. This is done based on the Debye-Bueche-Brinkman representation of the solution as a permeable medium where the obstacles against the fluid flow are the polymer coils themselves.

**THE HYDRODYNAMIC ROUSE-ZIMM MODEL OF POLYMER DYNAMICS**

Within the RZ model the polymer is represented by a chain of $N$ beads undergoing in solution the Brownian motion. The motion of the $n$th polymer bead is described by the equation

$$M \frac{d^2 \vec{x}_n(t)}{dt^2} = \vec{f}_n^{fr} + \vec{f}_n^{ch} + \vec{f}_n. \tag{1}$$

Here $\vec{x}_n$ is the position vector of the bead, $M$ is the mass of the bead, $\vec{f}_n^{ch}$ is the force with which the neighboring beads act on the $n$th bead, $\vec{f}_n$ is the random force due to the motion of the molecules of solvent, and $\vec{f}_n^{fr}$ is the friction force on the bead during its motion in the solvent. In the original theory with the hydrodynamic interaction taken into account the latter force was the Stokes one [1, 2],

$$\vec{f}_n^{fr} = -\xi \left[ \frac{d\vec{x}_n}{dt} - \vec{v}(\vec{x}_n) \right], \tag{2}$$

with $\vec{v}(\vec{x}_n)$ being the velocity of the solvent in the place of the $n$th bead due to the motion of other beads. The friction coefficient is $\xi = 6\pi\eta b$. As discussed in Introduction, this expression holds only for the steady-state flow. In a more general case the force (2) should be replaced by the Boussinesq force which, for any of the Cartesian components, has the form [16 - 18]

$$f^{fr}(t) = -\xi v(t) - \frac{1}{2} M_s \dot{v}(t) - 6\sqrt{\frac{M_s \xi}{2\pi}} \int_0^\infty \dot{v}(t - \beta^2) d\beta, \qquad (3)$$

where $M_s$ is the mass of the solvent displaced by one bead. Equations (1 – 3) have to be solved together with the hydrodynamic (Navier-Stokes and continuity) equations for the macroscopic velocity of the solvent,

$$\rho \frac{\partial \vec{v}}{\partial t} = -\nabla p + \eta \Delta \vec{v} + \vec{\varphi}, \quad \operatorname{div} \vec{v} = 0. \qquad (4)$$

Here $p$ is the pressure. The quantity $\vec{\varphi}$ has the sense of an external force per unit volume [1, 2],

$$\vec{\varphi}(\vec{x}) = -\sum_n \vec{f}_n^{fr}(\vec{x}_n) \delta(\vec{x} - \vec{x}_n). \qquad (5)$$

The solution of the hydrodynamic equations (4) can be, for any of the component $\alpha$ ($x$, $y$, or $z$), written in the form [26 - 29]

$$v_\alpha^\omega(\vec{r}) = \int d\vec{r}' \sum_\beta H_{\alpha\beta}^\omega(\vec{r} - \vec{r}') \varphi_\beta^\omega(\vec{r}'), \qquad (6)$$

where

$$H_{\alpha\beta}^\omega(\vec{r}) = A \delta_{\alpha\beta} + B \frac{r_\alpha r_\beta}{r^2}, \qquad (7)$$

$$A = \frac{1}{8\pi\eta r}\left[e^{-y} - y\left(\frac{1-e^{-y}}{y}\right)''\right], \quad B = \frac{1}{8\pi\eta r}\left[e^{-y} + 3y\left(\frac{1-e^{-y}}{y}\right)''\right]$$

plays a role of the Oseen tensor [1, 2] but in the Fourier representation. We have denoted here $y = r\chi$, $\chi = \sqrt{-i\omega\rho/\eta}$ (Re$\chi > 0$), and the prime stays for the differentiation with respect to $y$. Now we have to find the Fourier transform (FT) of Eq. (5), and then $\varphi_\beta^\omega$ substitute to $v_\alpha^\omega$ from Eq. (6) and the result into the FT of equation of motion (1), with the use of the Boussinesq force (3). This results in the generalization of the RZ equation, which has the following form:

$$-i\omega x_{n\alpha}^\omega = \frac{1}{\xi^\omega}\left(f_{n\alpha}^{ch,\omega} + f_{n\alpha}^\omega + M\omega^2 x_{n\alpha}^\omega\right) + \sum_\beta \sum_{m \neq n} H_{\alpha\beta}^\omega(\vec{x}_n - \vec{x}_m)\left[f_{m\beta}^{ch,\omega} + f_{m\beta}^\omega + M\omega^2 x_{m\beta}^\omega\right]. \qquad (8)$$

Here $\xi^\omega = \xi\left[1 + \chi b + (\chi b)^2/9\right]$. For long chains the continuum approximation is used [1, 2]. Then the effective force between the beads (it follows from their equilibrium (Gaussian) distribution [2]) is $\vec{f}_n^{ch} \to 3k_B T a^{-2} \partial^2 \vec{x}(t,n)/\partial n^2$, with $a$ being the mean square distance between neighboring beads along the chain. The obtained equation (8) is nonlinear. As in the original model, we solve it using the preaveraging of the Oseen tensor [1, 2]. The tensor is replaced with its average over the equilibrium distribution $P(r_{nm})$,

$$P(r_{nm}) = (2\pi a^2 |n-m|/3)^{-3/2} \exp[-3r_{nm}^2/(2a^2|n-m|)], \quad \vec{r}_{nm} \equiv \vec{x}(n) - \vec{x}(m),$$

$$\left\langle H_{\alpha\beta nm}^{\omega} \right\rangle_0 = \left\langle A(r_{nm})\delta_{\alpha\beta} + B(r_{nm})\frac{r_{nm\alpha}r_{nm\beta}}{r_{nm}^2} \right\rangle = \delta_{\alpha\beta} h^{\omega}(n-m). \quad \vec{r}_{nm} \equiv \vec{x}_n - \vec{x}_m \tag{9}$$

$$h^{\omega}(n-m) = (6\pi^3|n-m|)^{-1/2} (\eta a)^{-1} [1 - \sqrt{\pi} z \exp(z^2) \operatorname{erfc}(z)], \quad z \equiv \chi a \sqrt{\frac{|n-m|}{6}}.$$

Now Eq. (8) contains only diagonal terms,

$$-i\omega \vec{x}^{\omega}(n) = \frac{1}{\xi^{\omega}} \left[ \frac{3k_B T}{a^2} \frac{\partial^2 \vec{x}^{\omega}(n)}{\partial n^2} + M\omega^2 \vec{x}^{\omega}(n) + \vec{f}^{\omega}(n) \right]$$

$$+ \int_0^N dm\, h^{\omega}(n-m) \left[ \frac{3k_B T}{a^2} \frac{\partial^2 \vec{x}^{\omega}(m)}{\partial m^2} + M\omega^2 \vec{x}^{\omega}(m) + \vec{f}^{\omega}(m) \right]. \tag{10}$$

This equation is easily solved with the help of the FT in the variable $n$, taking into account the boundary conditions at the ends of the chain [2]: $\vec{x}^{\omega}(n) = \vec{y}_0^{\omega} + 2\sum_{p=1}^{\infty} \vec{y}_p^{\omega} \cos(\pi n p / N)$, $\partial \vec{x}(t,n)/\partial n = 0$ at $n = 0$ and $n = N$. The inverse FT yields the following equation for the Fourier components $\vec{y}_p^{\omega}$:

$$\vec{y}_p^{\omega} = \vec{f}_p^{\omega} \left[ -i\omega \Xi_p^{\omega} - M\omega^2 + K_p \right]^{-1}, \tag{11}$$

where we have denoted

$$\Xi_p^{\omega} = \xi^{\omega} \left[ 1 + (2 - \delta_{p0}) N \xi^{\omega} h_{pp}^{\omega} \right]^{-1}, \quad K_p = \frac{3\pi^2 k_B T}{N^2 a^2} p^2, \quad p = 0, 1, 2\ldots \tag{12}$$

The matrix $h_{pp}^{\omega}$ is defined by the expression

$$h_{pq}^{\omega} = \frac{1}{N^2} \int_0^N dn \int_0^N dm\, h^{\omega}(n-m) \cos\frac{\pi p n}{N} \cos\frac{\pi p m}{N}. \tag{13}$$

In obtaining Eq. (11) we have already taken into account that the nondiagonal elements of the matrix are small in comparison with the diagonal ones and can be in the first approximation neglected. This "diagonalization approximation" has been proven in the case without memory [1, 2]. In our case it should be substantiated by numerical calculations. However, since we are interested in the long-time properties of the chain, where only small corrections to the classical results are expected, the approximation is reasonable. The obtained equation (11) can be investigated as it is usually done in the theory of the Brownian motion using the fluctuation-dissipation theorem (FDT) [19, 31] or the correlation properties of the forces $\vec{f}_p^{\omega}$ [22, 27 - 29],

$$\left\langle f_{p\alpha}^{\omega} f_{q\alpha}^{\omega'} \right\rangle = \frac{k_B T}{(2-\delta_{p0})\pi N} \operatorname{Re} \Xi_p^{\omega} \delta_{\alpha\beta} \delta_{pq} \delta(\omega+\omega'). \tag{14}$$

Equation (11) then yields the following expression for the time correlation function of the Fourier components $y_{\alpha p}(t)$:

$$\psi_p(t) = \left\langle y_{\alpha p}(0) y_{\alpha p}(t) \right\rangle = \frac{k_B T}{(2-\delta_{p0})\pi N} \int_{-\infty}^{\infty} d\omega \cos\omega t \frac{\operatorname{Re} \Xi_p^{\omega}}{\left|-i\omega\Xi_p^{\omega} - M\omega^2 + K_p\right|^2}, \tag{15}$$

in agreement with the FDT [31]. Another form of Eq. (15) is

$$\psi_p(t) = \frac{k_B T}{\pi i} \int_{-\infty}^{\infty} \frac{d\omega}{\omega} \alpha_p(\omega) \cos\omega t, \tag{15a}$$

where $\alpha_p(\omega)$ is a generalized susceptibility. The use of the Kramers-Kronig dispersion relation [31] immediately gives the initial value $\psi_p(0) = k_B T (2NK_p)^{-1}$, $p > 0$. Equation (15) represents the solution of the hydrodynamic RZ model, within the approximations described, for the Fourier amplitudes of the correlation functions of the positions of beads. Knowing $\psi_p(t)$, other correlation functions of interest can be found from Eq. (15), e.g. the velocity autocorrelation function (VAF) and the MSD [22]:

$$\phi_p(t) = \left\langle v_{\alpha p}(0) v_{\alpha p}(t) \right\rangle = -\frac{d^2 \psi_p(t)}{dt^2}, \qquad \left\langle \Delta y_p^2(t) \right\rangle = 2[\psi_p(0) - \psi_p(t)]. \tag{16}$$

## ANALYTICAL SOLUTIONS OF THE HYDRODYNAMIC RZ MODEL

*Steady-state limit*

The results of the RZ model in the case when the inertia effects are not considered follow from Eq. (15) if $\omega = 0$ is put in Eq. (12) and $M = 0$. For the mode $p = 0$ that describes the motion of the center of inertia of the coil [1, 2] we have

$$\psi_0(0) - \psi_0(t) = Dt, \tag{17}$$

with the diffusion coefficient

$$D = D_R + D_Z, \qquad D_R = \frac{k_B T}{6\pi N b \eta}, \qquad D_Z = k_B T h_{00}^0 = \frac{8 k_B T}{3(6\pi^3 N)^{1/2} a \eta} \tag{18}$$

(the indices *R* and *Z* stay for the Rouse and Zimm limits). The internal modes ($p \neq 0$) relax exponentially,

$$\psi_p(t) = \frac{k_B T}{2NK_p} \exp(-|t|/\tau_p), \tag{19}$$

with the relaxation times $\tau_p$. The relaxation rates can be expressed in the form

$$\frac{1}{\tau_p} = \frac{1}{\tau_{pR}} + \frac{1}{\tau_{pZ}}, \qquad \tau_{pR} = \frac{2N^2 a^2 b \eta}{\pi k_B T p^2}, \qquad \tau_{pZ} = \frac{\left(\sqrt{N}a\right)^3 \eta}{\left(3\pi p^3\right)^{1/2} k_B T}. \tag{20}$$

Usually only the limiting (Rouse or Zimm) cases are considered. The applicability of a specific model is controlled by the so-called draining parameter $h = 2\sqrt{3N/\pi} b/a$ that indicates whether the hydrodynamic interaction in solution is strong ($h \gg 1$, then the dynamics is of the Zimm type) or not ($h \ll 1$, the Rouse dynamics). For the diffusion of the coil as a whole the corresponding "draining parameter" is $D_Z / D_R = 4\sqrt{2}h/3$. For the internal modes the draining parameter depends on the mode number *p*: $h(p) = \tau_{pR}/\tau_{pZ} = h/\sqrt{p}$. Due to this, beginning from some *p*, all the higher internal modes become the Rouse modes. In general, the polymer is not a pure Zimm or Rouse one; the polymer dynamics should always possess features of both the models. This should be taken into account (but is often neglected) in the determination of the phenomenological model parameters (*N*, *a*, *b*) from experiments. For example, in recent experiments [32] the motion of the individual monomer within the polymer coils has been observed for the first time using the fluorescence correlation spectroscopy. The measured MSD of the monomer was fitted by the Rouse and Zimm limits assuming continuous distribution of the internal relaxation modes. In this approximation the MSD is proportional to $\sqrt{t}$ and $t^{2/3}$ in the Rouse and Zimm case, respectively [9, 10]. It was concluded that the double-stranded DNA surprisingly follows the Rouse-type dynamics in spite of the common belief that the behavior of this (although semiflexible) polymer is close to the Zimm-type dynamics. However, coming from the "joint" RZ theory described above, with the discrete internal modes, it can be shown that this conclusion is misleading and the long DNA's behave predominantly as the Zimm polymers [33, 34].

When the memory effects are considered, the following analytical results can be obtained.

*The Rouse limit*

This limit assumes that the hydrodynamic interaction contribution to the quantity $\Xi_p^\omega$ from Eq. (12) is negligible for all $\omega$. Accordingly, the subsequent equations change only by the substitution $\Xi_p^\omega \approx \xi^\omega$. This case is considered in more detail in our work [27]. To get analytical results for the motion of the whole coil (*p* = 0), in the Rouse limit one can adapt the results of the hydrodynamic theory of the Brownian motion of one particle [18] (usually the work [35] is cited). In the Rouse model the friction force on the coil is just a sum of the forces on individual beads (since the solvent is nonmoving). The results for one bead and that for the whole coil thus differ only by the factor 1/*N*. For example, the VAF of the coil is

$$\phi_0(t) = \frac{\phi_0(0)}{\lambda_1 - \lambda_2} \sum_{i=1,2} (-1)^{i+1} \lambda_i \exp(\lambda_i^2 t) \operatorname{erfc}(\lambda_i \sqrt{t}), \qquad (21)$$

where $\phi_0(0) = k_B T / (M + M_s/2) N$, $\lambda_i$ are the complex roots of the equation $\lambda^2 + \sqrt{\tau_b} \lambda / \tau + 1/\lambda = 0$ with $\tau = (M + M_s/2)/\xi$, and $\tau_b = b^2 \rho/\eta$ defines the time for passing the distance $b$ by the shear wave. At long times $t \gg \tau_b$ Eq. (21) yields the following asymptotic expression:

$$\langle \Delta y_0^2(t) \rangle = 2 D_R t \left[ 1 - \frac{2}{\sqrt{\pi}} \left( \frac{\tau_b}{t} \right)^{1/2} + \frac{2}{9} \left( 4 - \frac{M}{M_s} \right) \frac{\tau_b}{t} - \frac{1}{9\sqrt{\pi}} \left( 7 - 4\frac{M}{M_s} \right) \left( \frac{\tau_b}{t} \right)^{3/2} + \ldots \right], \qquad (22)$$

For short times we have

$$\langle \Delta y_0^2(t) \rangle \approx \frac{k_B T}{N(M + M_s/2)} t^2, \qquad (23)$$

where, however, $t \gg b/c$ ($c$ is the velocity of sound); the condition is due to the fluid incompressibility. The physically correct limiting value $k_B T t^2 / NM$ can be obtained only if the compressibility is taken into account, as discussed already in Ref. [18] (for a detailed solution of this "hydrodynamic paradox" see Ref. [22]).

In the case $p \neq 0$ ($K_p \neq 0$) the integral in Eq. (15) can be represented through the inverse Laplace transform [22]. The integrand is then expanded into a sum of elementary fractions, for which the Laplace transforms are known. This allows one to express the searched integral in a closed form through the error functions. The asymptotic expansions are then derived using known properties of these functions. The long-time asymptote for the function $\psi_p(t)$ has the form

$$\psi_p(0) - \psi_p(t) = \frac{k_B T}{2 N K_p} \left\{ 1 + \frac{1}{2\sqrt{\pi}} \frac{\tau_{pR} \tau_b^{1/2}}{t^{3/2}} \left[ 1 + 3 \frac{\tau_{pR}}{t} + \ldots \right] \right\}. \qquad (24)$$

Comparing these formulas with the results of the Rouse theory, it is seen that for the internal modes ($p \geq 1$) the value $\psi_p(0) = k_B T/(2NK_p)$ is the same as in the original model. The time dependence of the MSD of the whole coil at short times and the function $\psi_p(t)$ are different, now $\sim t^2$. The diffusion coefficient of the coil, $D_R$, is the Rouse one. At long times, in addition to the Einstein term in the MSD, there are other contributions, the leading of which is $\sim t^{1/2}$. A difference reveals also in the long-time dependence of other time correlation functions of the coordinates and velocities of the polymer segments in the Fourier representation. The inclusion of the hydrodynamic memory leads to the relaxation of the internal modes that differs from the traditional exponential decay of the correlation functions $\psi_p(t)$ and $\phi_p(t)$. This is reflected in the long-time tails of these functions. More than three decades ago the discovery of the tails of the molecular VAF in simple liquids (first by means of computer experiments [36, 37]) has led to an enormous number of investigations in the field of the statistical theory of liquids and in the theory of Brownian motion [22]. We believe that similar peculiarities found here could stimulate new studies in the dynamics of polymers, first of all by means of computer simulations.

As an illustration of the difference between the original Rouse model and the model with hydrodynamic memory we show on Fig. 1 the time dependence of the VAF of the center of inertia of the coil, calculated from Eqs. (15) and (16).

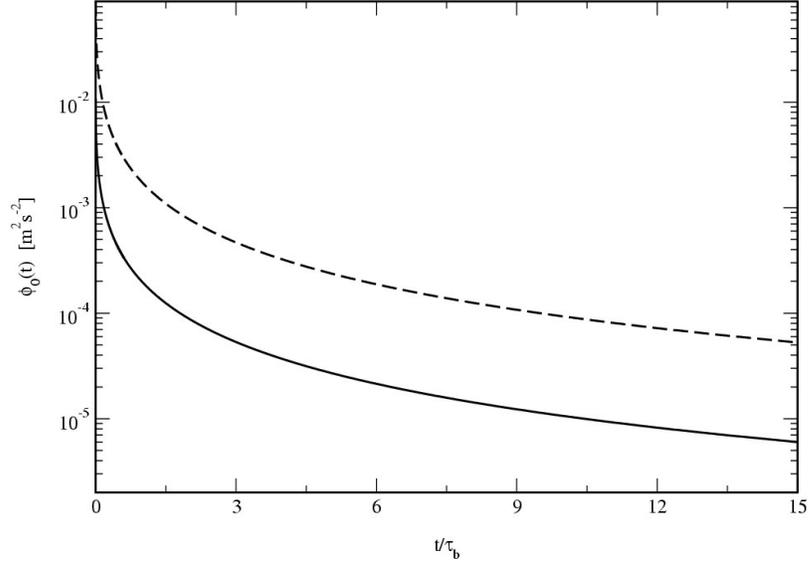

**Fig. 1** Velocity autocorrelation function for the center of inertia of the coil, calculated in the Rouse limit of Eq. (15). In the original Rouse theory $\phi_0(t) = 0$. The solvent and polymer parameters are taken from Ref. [32] and correspond to dsDNA in water environment: $T = 293$ K, $\eta = 10^{-3}$ Pa·s, $a = 10^{-7}$ m, $b = 12.8 \cdot 10^{-9}$ m, $M = 3.2 \cdot 10^{-22}$ kg. The upper curve is for $N = 9$, the lower one for $N = 79$ (2400 and 23100 base pairs DNA, respectively).

*The Zimm limit*

When the hydrodynamic interaction is strong for all frequencies that significantly contribute to the studied correlation functions, we have from Eq. (12)

$$\Xi_p^\omega \approx \left[(2 - \delta_{p0}) N h_{pp}^\omega\right]^{-1}. \qquad (25)$$

However, the inertial effects of the viscous solvent are still included into the consideration. The Oseen matrix (13) can be calculated with any degree of precision, e.g. for $p = 0$ we have the exact result

$$h_{00}^\omega = \frac{2}{\sqrt{6N}\pi\eta a}\frac{1}{\chi_\omega}\left[1 - \frac{2}{\sqrt{\pi}\chi_\omega} + \frac{1}{\chi_\omega^2}\left(1 - \exp\chi_\omega^2 \operatorname{erfc}\chi_\omega\right)\right], \quad \chi_\omega \equiv \sqrt{\frac{N}{6}}\chi a. \qquad (26)$$

When $p \neq 0$, one can act similarly as in Ref. [2] and calculate the Oseen matrix (13) using the rapid decrease of $h^\omega(n-m)$ from Eq. (9) with the increase of $|n-m|$ ($h^\omega(n-m) \sim |n-m|^{-3/2}$ at large $|n-m|$, i.e. more rapidly than in the original Zimm theory where $h(n-m) \sim |n-m|^{-1/2}$ at $\omega = 0$). The nondiagonal components of the matrix are small, and for the diagonal ones we found [26]

$$h_{pp}^{\omega} \approx \frac{1}{\pi \eta \, a \sqrt{3\pi N p}} \frac{1 + \chi_p}{1 + (1 + \chi_p)^2}, \qquad \chi_p = \sqrt{\frac{N}{3\pi p}} \chi a. \tag{27}$$

This result will be used later when the influence of other polymers in solution on the dynamics of our chosen "test" polymer will be discussed. The correlation functions in the *t*-representation can be now obtained by standard methods [22]. Their long-time asymptotes are as follows. In the case of the diffusion of the coil as a whole we obtain

$$\psi_0(0) - \psi_0(t) \approx D_Z \left[ t - \left(\frac{3N\rho}{32\eta}\right)^{1/2} t^{1/2} + \ldots \right] = D_Z \left[ t - \frac{2}{\sqrt{\pi}} (\tau_R t)^{1/2} + \ldots \right], \tag{28}$$

where the Zimm diffusion coefficient $D_Z$ is determined in Eq. (18). The second equality has exactly the form familiar in the theory of the Brownian motion of rigid particles of radius $R$. Now the characteristic time $\tau_R = R^2 \rho / \eta$ is expressed through the hydrodynamic radius of the Zimm coil, determined from the relation $D_Z = k_B T/(6\pi R \eta)$. In the standard model of Gaussian chains the radius $R$ coincides with the Kirkwood expression [1]

$$R^{-1} = N^{-2} \left\langle \sum_{n=1}^{N} \sum_{m=1, m \neq n}^{N} r_{nm}^{-1} \right\rangle \approx \sqrt{\frac{2}{3\pi N}} \frac{8}{a}, \tag{29}$$

valid for $N \gg 1$, when the summation can be replaced by integration. Consider now the internal modes of the polymer. The components of the Oseen matrix are known to any desired power of $(-i\omega)^{1/2}$. In the lowest approximation for long times we have obtained for the correlation function $\psi_p(t)$ [28, 29],

$$\frac{\psi_p(t)}{\psi_p(0)} \approx -\frac{2^9}{45\pi^3} \sqrt{\frac{2}{\pi}} \left( 1 + \frac{16}{3\pi^2 p} \frac{\tau_R}{\tau_{pZ}} \right) \frac{1}{p^3} \frac{\tau_{pZ} \tau_R^{3/2}}{t^{5/2}}, \tag{30}$$

where $p \geq 1$, and $\tau_{pZ}$ is the Zimm relaxation time (20).

Below are examples of the numerical calculations of relevant correlation functions describing the polymer motion. The calculations were done from Eq. (15) with no assumption on the validity of a specific, Rouse or Zimm, limit. Figures 2 - 5 illustrate the peculiarities of the polymer dynamics with the inertia effects included into the consideration.

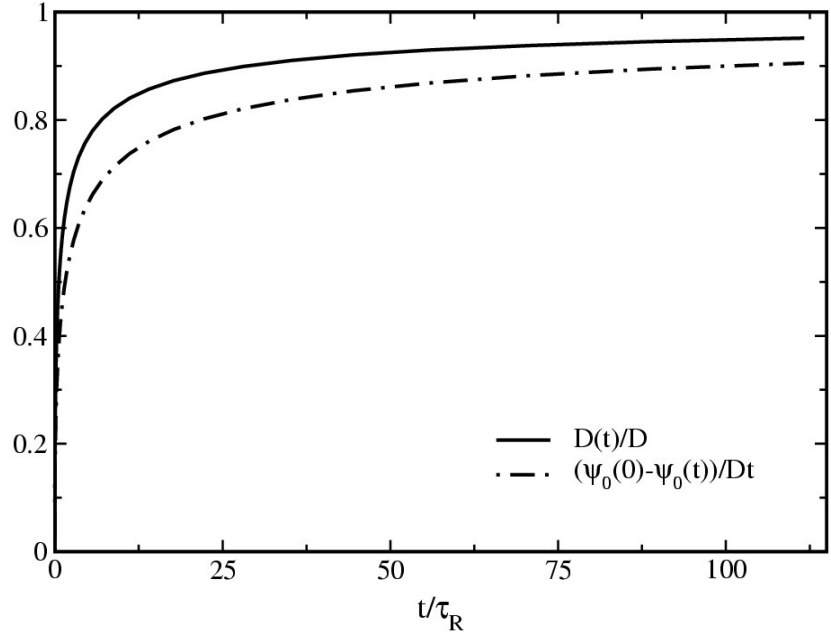

**Fig. 2** Illustration of the difference between the original RZ theory and our results represented by the dependence of the correlation functions $\Psi_0(0) - \Psi_0(t)$ describing the motion of the polymer as a whole and $D(t)$ (the time-dependent diffusion coefficient defined as $-d\psi_0(t)/dt$) on the dimensionless time $t/\tau_R$. The functions calculated from Eq. (15) are related to the Einstein limits $Dt$ and $D$ from the previous RZ theory. The parameters are for the long double-stranded DNA studied in Ref. [32] ($a = 10^{-7}$ m, $b = 12.8 \cdot 10^{-9}$ m, $\rho = 10^3$ kg/m$^3$, $\eta = 1$ mPa·s) except the DNA length ($4 \cdot 10^4$ base pairs, $N = 1360$), giving $\tau_R = 1$ μs.

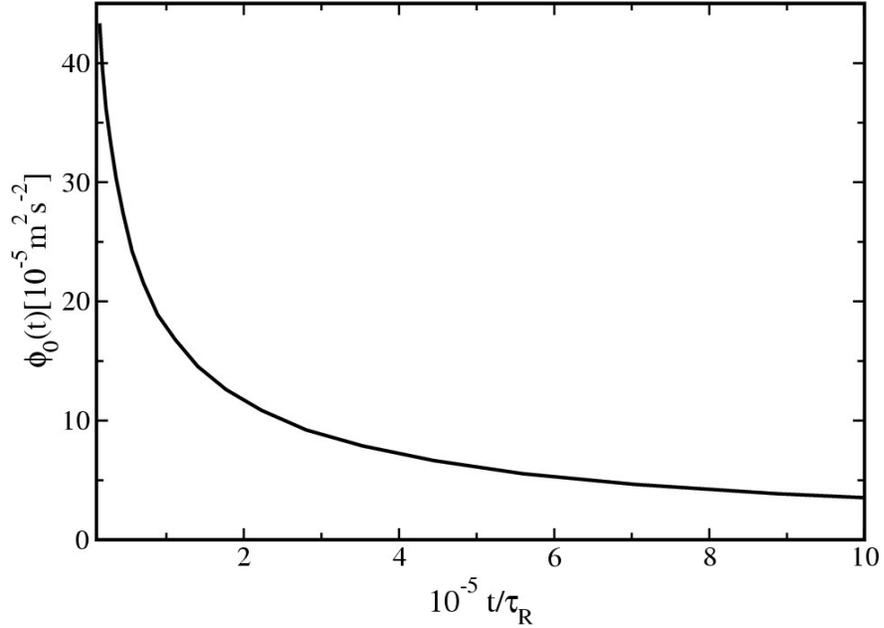

**Fig. 3.** Velocity autocorrelation function for the motion of the center of inertia of the polymer coil in the model with hydrodynamic memory. The parameters are the same as in Fig. 2. In the RZ theory without the inertia effects taken into account $\phi_0(t) = 0$.

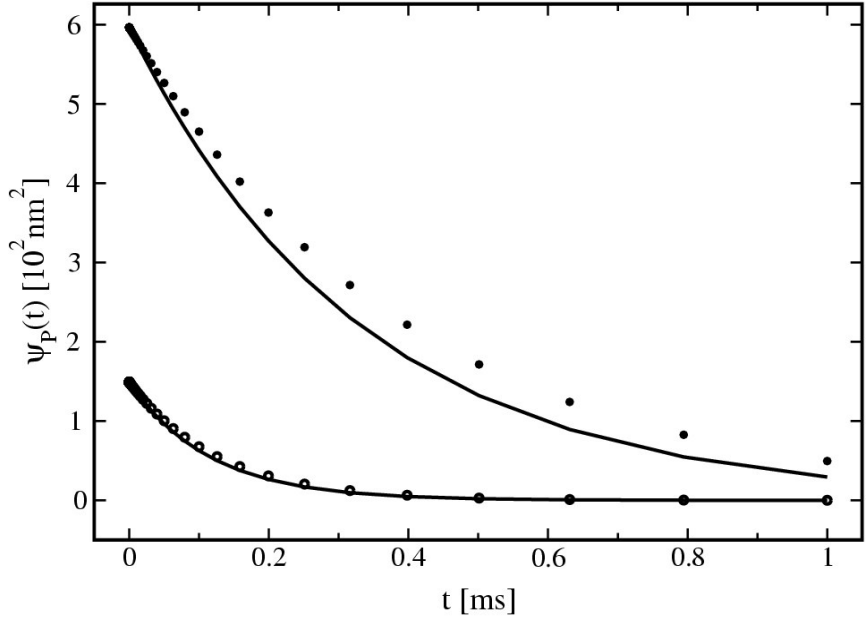

**Fig. 4** Correlation functions $\psi_p(t)$ calculated from Eq. (15). The filled circles correspond to the lowest internal mode $p = 1$, the open ones are for $p = 3$. In the original theory $\psi_p(t)$ relaxed exponentially (lines). The parameters for the single stranded DNA (6700 bases) have been obtained by us optimizing the RZ theory to the data from Ref. [32] ($N = 422$, $a = 9.15$ nm, $b = 4.56$ nm). The solvent characteristics were $\eta = 0.69$ mPa·s, $\rho = 10^3$ kg/m$^3$, and $T = 310$ K.

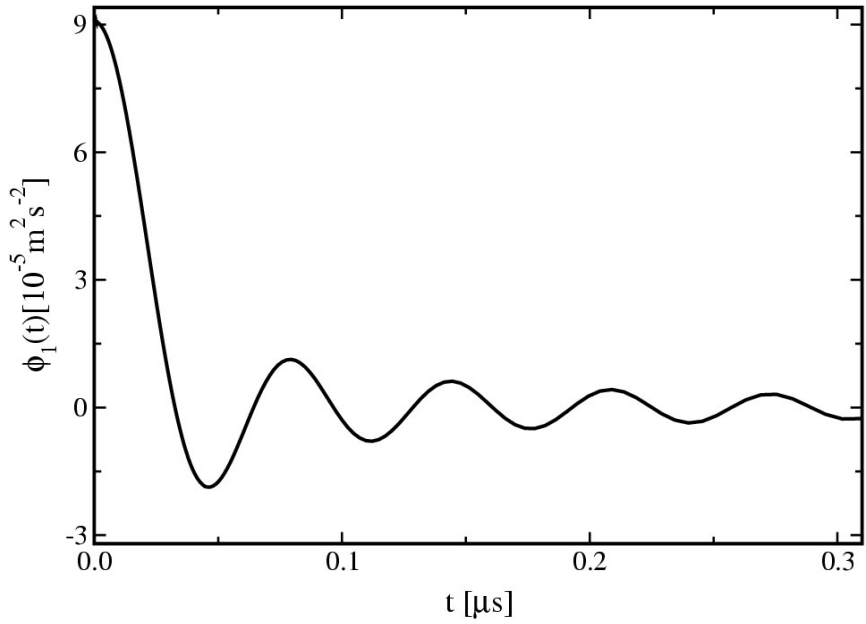

**Fig. 5** Velocity autocorrelation function for the $p = 1$ internal mode and the same parameters as in Fig. 4. In the previous RZ theory $\phi_1(t) < 0$ and relaxed exponentially. The corresponding curve is on the graph indistinguishable from the zero value.

**TIME SCALES IN THE POLYMER MOTION**

The motion of an individual polymer in solution, from its microscopic origin to macroscopically observable phenomena, can be characterized by several very different time and length scales. The relevant times for the Rouse polymer are seen from Eq. (15) if we express it in the form

$$\psi_p(t) = \frac{k_B T \tau_p^2}{2\pi N \xi} \int_{-\infty}^{\infty} d\omega \cos\omega t \frac{1 + \mathrm{Re}(-i\omega\tau_b)^{1/2}}{\left|1 - i\omega\tau_p\left[1 + (-i\omega\tau_b)^{1/2} - i\omega\tau_b/9\right] - \tau_B\tau_p\omega^2\right|^2}. \tag{15a}$$

Here

$$\tau_B = \frac{M}{\xi} = \frac{2}{9}\frac{\rho_B}{\rho}\tau_b \tag{31}$$

is the characteristic time in the Ornstein theory of the Brownian motion (with the Stokes force as the friction force on the particle). This time determines the scale of exponential relaxation of the free and massive Brownian particle of radius $b$; $\rho_B/\rho$ is the ratio between the density of the particle and the solvent. The nonexponential relaxation of the VAF is characterized by the time $\tau_b$. Only when $\rho_B \gg \rho$ the Ornstein description of the Brownian motion is adequate, as first discussed in Ref. [18]. For the usual case of neutrally buoyant particles both the times are comparable and the effects of fluid inertia should be taken into account at such time scales. Another characteristic time in Eq. (15a) is $\tau_p = \xi/K_p$. For one particle the quantity $K_p$ plays a role of the force constant of an external harmonic field [22], in our case it is determined in Eq. (12). In addition to the times entering Eq. (15a), the sound traversal time is $\tau_s = b/c$, where $c$ is the speed of sound (~ 1500 m/s in water at room conditions). For particles with $b \sim 100$ nm, $\tau_s$ is less than $10^{-10}$ s. The characteristic time of diffusion $\tau_D = b^2/D$ (the time during which the particle moves on the distance $b$) for such particles is of order of milliseconds. Finally, the particle is subject to interaction with the molecules of solvent. The time scale characterizing their collisions with the Brownian particle is $\tau_c \sim 10^{-12}$ s. For large Brownian particles ($b \gg$ the distance between the molecules) and the times much longer than $\tau_c$ the collisions can be considered as a stochastic noise and the solvent as a continuum, and described by the Navier-Stokes equations. The vorticity time $\tau_b \gg \tau_c$ (for $b \sim 100$ nm $\tau_b \sim 10^{-8}$ s). In the bead-spring models of polymers the bead radius is a phenomenological parameter, which however should be much smaller than the size of the whole polymer. For nanometer-sized beads the times $\tau_b$ and $\tau_c$ are comparable so that the values of the random force on the bead cannot be considered as statistically independent (the Langevin force delta-correlated in the time) at the time scale $\tau_b$. Our model in its full formulation thus requires $b$ much larger than the atomic dimensions, and $\tau_b \gg \tau_c$. If the last condition is not satisfied, the model applies to the times $t \gg \tau_b$ and $\tau_c$. This means that instead of the Boussinesq force we again use the Stokes force as in the classical theory. Due to this the memory effects will play no role in the Rouse limit. However, they are still present in the general case. It is well seen in the Zimm limit when the nondiffusive motion of the polymer is characterized by the time scale $\tau_R$ determined by the radius of the whole polymer $R$. The simplest variant of our theory thus consists in replacing $\xi^\omega$ (see Eq. (8)) by $\xi$ so that $\Xi_p^\omega$ in Eq. (12), (14) and (15) becomes $\Xi_p^\omega = \xi[1 + (2 - \delta_{p0})N\xi h_{pp}^\omega]^{-1}$.

Using the above characteristic time scales one can classify the regimes in the Brownian motion or in the polymer dynamics and judge about their importance for the observed quantities. For example, the known Ehrenfest's paradox [18] (the initial value of the VAF in incompressible fluid is in disagreement with the equipartition theorem) is explained as follows. In the time scale $\tau_c$ the VAF of the Brownian particle decays from the equipartition value $k_B T/M$ to $k_B T/(M + M_s/2)$ through emission of sound waves. It is interesting that for longer times the VAF not only decays but is an oscillating function of time. Such a behavior was discovered in computer experiments for a single molecule in a liquid [36] and can be qualitatively explained as a result of the collective elastic reaction of the surrounding molecules to the motion of the particle [38]. A similar relaxation occurs for the VAF $\phi_p(t)$ of the internal modes of our polymer as a result of the effective interaction between the beads (see Fig. 5). Note that in the previous model $\phi_p(t) < 0$ and monotonically converges to zero.

Let us return to Eq. (15a) for the correlation functions of the Rouse polymers and estimate how the decay of these functions differs from the exponential relaxation in the classical model. We have the estimation

$$\frac{\tau_b}{\tau_1} \sim \frac{\pi k_B T \rho}{2\eta^2} \frac{b}{N^2 a^2}. \tag{32}$$

Taking the extreme value for the bead radius, $b \sim a/2$ (when the beads are close to one another), one obtains $\tau_b/\tau_1 \sim \pi k_B T\rho/8\eta NR$. For a typical hydrodynamic radius $R \sim 100$ nm, when the solvent is water at room conditions, $\tau_b/\tau_1 < 2\times 10^{-5}/N$. In the Zimm limit we would have

$$\frac{\tau_R}{\tau_1} = \frac{(3\pi)^2}{2^{10}\sqrt{2}} \frac{k_B T\rho}{\eta^2 R} \sim 0.06 \frac{k_B T\rho}{\eta^2 R}, \tag{33}$$

i.e., for the same conditions $\tau_R/\tau_1 \sim 2.5\times 10^{-6}$. It is seen from Eq. (15a) that due to the smallness of $b$ the frequency $\omega$ must become very large to have the second term in the denominator comparable to 1 and the expression in |...| is $\approx 1 - \omega\tau_p(i + \omega\tau_B)$ up to relatively high frequencies. Since $\tau_B$ is also small, the integrand is determined mainly by the expression $|1 - i\omega\tau_p|^{-2}\cos\omega t$ so that the result of integration is $\sim \exp(-t/\tau_p)$. The relaxation of the internal Rouse modes is thus very close to exponential. The frequencies at which the terms with the characteristic times $\tau_b$ and $\tau_B$ become appreciable are so high that the contribution to the integral due to the rapidly oscillating function $\cos\omega t$ is small.

Analogous estimations for the Zimm limit can be done using the additional inequality $\tau_R/\tau_1 \ll 1$. In both the limits the most evident differences between the classical model and the model with memory occur at short times as illustrated by Figs. 1 - 3 for the motion of the whole polymer, and by Fig. 5 for the internal modes. In spite of the qualitative difference of the VAF, at experimentally accessible times the correlation functions $\psi_p(t)$ for the internal modes relax nearly exponentially, as demonstrated by Fig. 4. This will be used below in the calculation of the polymer dynamic structure factor. The MSD of the center of inertia of the coil is smaller than in the previous theory (Fig. 2). The shorter times of observation, the more significant is this difference.

**THE DYNAMIC STRUCTURE FACTOR**

The quasielastic scattering of light or neutrons from a polymer coil in dilute solution is described by the Van Hove function (the intermediate scattering function or the dynamic structure factor) of an individual coil [1, 2],

$$G(\vec{k},t) = \frac{1}{N} \sum_{m,n} \langle \exp\{i\vec{k}[\vec{x}_n(t) - \vec{x}_m(0)]\} \rangle, \quad t > 0. \tag{34}$$

Here $\vec{k}$ is the wave-vector change at the scattering and $\vec{x}_n(t)$ is the position of the $n$th bead at the time $t$. Since $\vec{x}_n(t) - \vec{x}_m(0)$ is a linear function of the Gaussian random force, the distribution of this quantity is also Gaussian [1],

$$\langle \exp\{i\vec{k}[\vec{x}_n(t) - \vec{x}_m(0)]\} \rangle = \exp\left\{-\frac{k^2}{2} \langle [x_n(t) - x_m(0)]^2 \rangle \right\}, \tag{35}$$

where $x_n$ is the projection of $\vec{x}_n$ on the vector $\vec{k}$. After the transition to the Fourier representation as after Eq. (10) we obtain

$$G(k,t) = \frac{1}{N} \sum_{nm} \exp[-k^2 \Phi_{nm}(t)], \tag{36}$$

where

$$\Phi_{nm}(t) = \psi_0(0) - \psi_0(t) + 2\sum_{p=1}^{\infty}\left[\psi_p(0)\left(\cos^2\frac{\pi np}{N} + \cos^2\frac{\pi mp}{N}\right) - 2\psi_p(t)\cos\frac{\pi np}{N}\cos\frac{\pi mp}{N}\right],$$

$$\psi_0(0) - \psi_0(t) = \frac{1}{2}\langle[y_0(t) - y_0(0)]^2\rangle, \qquad \psi_{p\geq 1} = \langle y_p(t) y_p(0) \rangle. \tag{37}$$

In the Rouse and Zimm models without the hydrodynamic memory the correlation functions $\psi_p$ describing the internal motion are exponential, $\psi_p \sim \exp(-t/\tau_p)]$, which is essentially used in the evaluation of the Van Hove function [1, 2]. The internal modes contribute practically only at large $kR$ ($R$ is the radius of the coil) and at small $\psi_p(0) - \psi_p(t)$, which implies $t \ll \tau_p$, since in other cases the Van Hove function is too small [1].
Consider now the situation when the hydrodynamic memory is taken into account.

*Long time approximation*
Although, as mentioned above, at long times the memory effects are small, we shall dwell on this case which is useful from the methodical point of view and also for the study of the kinetics of individual monomers, e.g. by fluorescence correlation spectroscopy (FCS) [32 - 34]. For the interpretation of the FCS measurements only the time dependence of the studied correlation functions is important - the measured mean square displacement of a monomer does not depend on the scattering vector $k$. Here this corresponds to the consideration of the time dependence of the function (37). The time behavior of the whole DSF, however, essentially depends on the region of the employed scattering vectors.

Taking into account that $\psi_p(0)$ from Eq. (15) is $\psi_p(0) = Na^2/6\pi^2 p^2$, Eq. (37) can be rewritten in the form

$$\Phi_{nm}(t) = R^2 \frac{t}{\tau_D}\left[1 - \frac{2}{\sqrt{\pi}}\left(\frac{\tau}{t}\right)^{1/2} + ...\right]$$
$$+ \frac{128}{(3\pi)^3} R^2 \sum_{p=1}^{\infty} \frac{1}{p^2}\left[\cos^2\frac{\pi np}{N} + \cos^2\frac{\pi mp}{N} - 2\frac{\psi_p(t)}{\psi_p(0)}\cos\frac{\pi np}{N}\cos\frac{\pi mp}{N}\right]. \qquad (38)$$

Here $\tau = \tau_b = b^2\rho/\eta$ for the Rouse model, in the Zimm case $\tau = \tau_R = R^2\rho/\eta$, and $\tau_D$ is the already discussed diffusion time of the whole coil. It is seen from Eq. (38) that at times $t \sim \tau_D$ both the terms on the right hand side are important. The first term begins to dominate for the times longer than $\tau_D$, and when $t \gg \tau_D$, the scattering is fully determined by the diffusion of the coil as a whole. This holds for *all values* of the scattering vectors. When $t \ll \tau_D$, the diffusion term becomes negligibly small and the contributions of internal modes prevail. Again, this is true independently on the values of the scattering vectors. The memory effects in the FCS thus could be revealed only at short times (if we are interested in the polymer internal motion) or at long times (for the motion of the whole polymer). However, since the characteristic time of the nondiffusive motion is negligibly small compared to the diffusion time, $\tau_R \ll \tau_D$, the hydrodynamic memory has no effect on the diffusion contribution to the MSD of an individual monomer. For the dynamic scattering experiments on the whole coil the consideration must be carried out more carefully. So, as will be shown below, the motion of the whole polymer can be studied also at short times, assuming the scattering vectors $k$ are small enough.

Let us return to the expression for the Van Hove function (36),

$$G(k,t) = \frac{1}{N}\exp\{-k^2[\psi_0(0) - \psi_0(t)]\}$$

$$\times \sum_{nm}\exp\left\{-\frac{Na^2k^2}{3\pi^2}\sum_{p=1}^{\infty}\frac{1}{p^2}\left[\cos^2\frac{\pi np}{N} + \cos^2\frac{\pi mp}{N} - 2\frac{\psi_p(t)}{\psi_p(0)}\cos\frac{\pi np}{N}\cos\frac{\pi mp}{N}\right]\right\}. \qquad (39)$$

The sums with cosines squared are known [39] so that for the sum in this expression we have

$$\sum_{mn}\exp\{...\} \approx \exp\left(-\frac{Na^2k^2}{9}\right)\sum_{mn}\exp\left[-\frac{Na^2k^2}{6}\left(\frac{m^2+n^2}{N^2} - \frac{m+n}{N}\right)\right]$$
$$\times \left[1 + \frac{2Na^2k^2}{3\pi^2}\sum_{p=1}^{\infty}\frac{1}{p^2}\frac{\psi_p(t)}{\psi_p(0)}\cos\frac{\pi np}{N}\cos\frac{\pi mp}{N}\right]. \qquad (40)$$

In the continuum approximation the sums over $m$ and $n$ are replaced by integrals. Using the fact that only small quantities $Na^2k^2/6 = R_G^2 k^2 \ll 1$, where $R_G$ is the gyration radius, are relevant for our consideration (in the opposite case the DSF becomes very small), after the integration we obtain

$$G(k,t) \approx N \exp\{-k^2[\psi_0(0)-\psi_0(t)]\}\exp\left(-\frac{Na^2k^2}{36}\right)$$
$$\times\left[1+\frac{2}{27}\left(\frac{Na^2k^2}{\pi^2}\right)^3 \sum_{p=1}^{\infty}\frac{\psi_p(t)}{\psi_p(0)}\frac{1}{p^6}\cos^2\frac{\pi p}{2}\right]. \qquad (41)$$

The lowest mode that contributes in this approximation is $p = 2$. Taking into account that for the Rouse and Zimm models the quantity $\psi_p(t)/\psi_p(0)$ depends on $p$ as $p^{-2}$ and $p^{-9/2}$, respectively, there is no reason to keep the higher modes, so that Eq. (41) can be rewritten in a more compact form,

$$G(k,t) \approx N \exp\{-k^2[\psi_0(0)-\psi_0(t)]\}\exp\left[-\frac{Na^2k^2}{36}\left(1-\frac{N^2a^4k^4}{24\pi^6}\frac{\psi_2(t)}{\psi_2(0)}\right)\right]. \qquad (42)$$

Substituting here the solutions for the corresponding correlation functions $\psi_p$ and the solution for the diffusion of the coil as a whole, we obtain the Van Hove function for the considered models. However, at the studied conditions not only the second exponential can be simply replaced by unity but also the first exponential practically does not differ (at the considered long times) from $\exp(-k^2 Dt)$, i.e. the memory effects could not be observed.

*Large scattering vectors*

Consider now large wave-vector transfers at the scattering, $kR_G \gg 1$. At this case the internal motion of the polymer is usually expected to be observed. As shown above, the internal modes relax practically exponentially, as in the classical models. The derivation of the DSF thus almost exactly repeats the calculations known from the literature [9, 10, 1]. For exponentially relaxing internal modes, $\psi_p(t)=\psi_p(0)\exp(-t/\tau_p)$, Eq. (37) can be rewritten as follows:

$$\Phi_{nm}(t) = \psi_0(0)-\psi_0(t)+\frac{a^2}{6}|m-n|$$
$$+\frac{2Na^2}{3\pi^2}\sum_{p=1}^{\infty}\frac{1}{p^2}\left(1-e^{-t/\tau_p}\right)\cos\frac{\pi mp}{N}\cos\frac{\pi np}{N}, \qquad (43)$$

where the second term comes from the sum [1]

$$\sum_{p=1}^{\infty}\frac{1}{p^2}\left(\cos\frac{\pi mp}{N}-\cos\frac{\pi np}{N}\right)^2 = \frac{\pi^2}{2N}|m-n|.$$

The consideration can be limited to the times $t \ll \tau_p$ since the DSF becomes very small in other cases [1]. The product of cosines in Eq. (43) is represented through a sum of cosines with the arguments $p\pi(n+m)/N$ and $p\pi(n-m)/N$. Since the sum is dominated by large $p$, for which the first cosine changes the sign very rapidly, its contribution becomes very small. The remaining term is evaluated by converting the sum over $p$ to the integral. Then in the DSF from Eq. (36) the sum over $n$ and $m$ is replaced by the double integral [1]. The approximate result for the Rouse case is

$$G(k,t) = \frac{12}{k^2 a^2} \exp\{-k^2[\psi_0(0) - \psi_0(t)]\} \int_0^\infty du \exp\left[-u - (\Gamma t)^{1/2} h\left(\frac{u}{(\Gamma t)^{1/2}}\right)\right], \qquad (44)$$

where

$$\Gamma = \frac{k_B T}{12\xi} k^4 a^2 = \frac{1}{2} D k^4 R_G^2. \qquad (45)$$

Here $\xi$ is the friction coefficient of one bead and the second equality is expressed through the diffusion coefficient of the Rouse coil and the gyration radius. In the previous theory $\Gamma$ was the first cumulant to the DSF. The function $h(u)$ is

$$h(u) = \frac{2}{\pi} \int_0^\infty dx \frac{\cos(xu)}{x^2} \left(1 - e^{-x^2}\right). \qquad (46)$$

When $\Gamma t \gg 1$ (for "short times" $t \ll \tau_p$ it is possible if $1/\Gamma \ll \tau_p$, i.e. if $6\pi k^2 a^2 \ll N/p$), the integral in Eq. (44) is further simplified to the stretched exponential $\sim \exp\left(-2\sqrt{\Gamma t/\pi}\right)$ [1].

In the Zimm case the DSF is obtained in the same way [10]. The quantity $\Gamma$ (the first cumulant in the Zimm model at large scattering vectors) is now

$$\Gamma = \frac{k_B T}{6\pi\eta} k^3 \qquad (47)$$

and

$$h(u) = \frac{2}{\pi} \int_0^\infty dx \frac{\cos(xu)}{x^2} \left[1 - \exp\left(-\frac{x^{3/2}}{\sqrt{2}}\right)\right]. \qquad (48)$$

For $\Gamma t \gg 1$ (which is possible if $\left(kN^{1/2} a / p^{1/2}\right)^3 \gg 60$, or $(kR_G)^3 \gg 4p^{3/2}$) the DSF in the Zimm model was

$$G(k,t) \approx G(k,0) \exp\left[-1.35(\Gamma t)^{2/3}\right]. \qquad (49)$$

For large scattering vectors the diffusion term in DSF is usually neglected [1]. In the case of Rouse model (for the solution in the form of the stretched exponential) it is possible when the time-dependent ratio

$$\zeta_R = \frac{k^2 Dt}{2(\Gamma t/\pi)^{1/2}} = \frac{(\pi \Gamma t)^{1/2}}{(kR_G)^2} = \frac{1}{p}\left(\frac{t}{\pi \tau_p}\right)^{1/2}$$

is much smaller than unity. Analogously, in the Zimm case the diffusion contribution in Eq. (49) can be neglected in comparison with the contribution from internal modes, if

$$\zeta_Z = \frac{1}{N^{1/2} a}\left(\frac{k_B T t}{\eta}\right)^{1/3} \approx \frac{0.7}{\sqrt{p}}\left(\frac{t}{\tau_p}\right)^{1/3} \ll 1.$$

It is seen from the last two expressions that even if $t/\tau_p$ is relatively small, the ratio between the diffusion contribution and the lowest internal mode can be notable (if $t/\tau_p \sim 10^{-2}$ in the Zimm case, the diffusion contribution differs only some 15% from the contribution due to the first internal mode). Note that one should be careful with the use of the elegant solutions in the form of the stretched exponentials [9, 10]. In fact, these are rather crude approximations valid in a restricted time interval. For instance, applying such solutions in the determination of the character of the monomer dynamics in the FCS experiments [32], the dynamics of the double-stranded DNA motion in water solution has been characterized as being of the Rouse type (the observed monomer MSD followed the $t^{1/2}$ law). However, a closer look at the validity of this approximation has revealed that the assumption of the continuous distribution of the internal modes (in the mode numbers $p$) fails for the experimental times. For the studied times the monomer MSD calculated as $\sim t^{1/2}$ was from 20% to one order larger than the MSD found using the correct (discrete in $p$) distribution of the polymer normal modes [34]. In such cases the identification of the polymer dynamics based on the $t^{1/2}$ ("Rouse") or $t^{2/3}$ ("Zimm") has obviously no value.

*Small scattering vectors*

In the classical monograph [1] this case is considered as follows. In the exponent in Eq. (36) only the diffusion term ($-k^2 Dt$) is considered and the other terms are neglected assuming that their magnitudes are smaller than $k^2 Na^2$. However, as seen from Eq. (38), all the terms in the exponential equally depend on $k$ and only the time dependence is essential for their comparison. In spite of this the result for $G(k,t) \sim \exp(-k^2 Dt)$ is correct. This can be shown by the direct calculation of the DSF, similarly as it was done above. For all range of $k$ these calculations are complicated but the desired result can be obtained by a simpler way, calculating the initial decay rate of the DSF [1]. In fact, since we are interested in the behavior of the DSF at short times where the influence of the memory is expected, the DSF can be approximated by the expression

$$G(k,t) = G(k,0)\exp(-\Gamma t). \tag{50}$$

The decay rate $\Gamma$ can be calculated rigorously [40, 41]. For $kR_G \ll 1$ the result is $\Gamma = k^2 D$ and $G(k,0) \to N$ as $k \to 0$. In the Zimm model $D$ in this expression is not the true long-time diffusion coefficient but the Kirkwood diffusion coefficient [1, 42]. Within the linearization approximation these coefficients are equal [1] and we will not distinguish them.

The only difference between our approach and the original theory (if the relaxation of the internal modes is assumed to be exponential) is the replacement of the factor $\exp(-k^2 Dt)$ by the exponential $\exp\{-k^2[\psi_0(0)-\psi_0(t)]\}$. Then the DSF at short times and small $k$ is expressed as

$$G(k,t) \approx G_{RZ}(k,0)\exp\{-k^2[\psi_0(0)-\psi_0(t)-Dt]-\Gamma t\}, \tag{51}$$

where $G_{RZ}$ and $\Gamma$ have the same sense as in the previous Rouse and Zimm theories. Below the first cumulant will be considered in more detail. Equation (51) can be used in estimations of the effects of hydrodynamic memory on the DSF at short times.

*Experimental time resolution and the first cumulant to DSF*

Every dynamic scattering experiment is characterized by the shortest accessible time. The existence of a finite resolution time has a significant role in the analysis of the data. As already mentioned in this work, there are discrepancies between the theoretical and experimental values of the first cumulant. According to the paper [13], the differences are caused by the principal underestimating of the first cumulant, which is due to the experimental resolution in the time. Let us assume that a measured time correlation function is a sum of exponential functions with the relaxation times $1/\Gamma_n$ and the amplitudes $A_n$. Then the observed first cumulant will be

$$\overline{\Gamma}_{\exp} = \sum_n A_n\Gamma_n \exp(-\Gamma_n\Delta\tau)\Big/\sum_n A_n\exp(-\Gamma_n\Delta\tau), \tag{52}$$

where $\Delta\tau$ is the time resolution. When $\Gamma_n\Delta\tau \ll 1$, after the expansion to the first order in $\Delta\tau$ we get

$$\overline{\Gamma}_{\exp} \approx \overline{\Gamma}\left[1+\frac{\overline{\Gamma^2}-\overline{\Gamma}^2}{\overline{\Gamma}^2}\overline{\Gamma}\Delta\tau\right], \tag{53}$$

where $\overline{\Gamma_n} \equiv \overline{\Gamma}$ and $\overline{\Gamma_n^2} \equiv \overline{\Gamma^2}$. Usually (e.g. for polydisperse systems) the difference $\overline{\Gamma^2}-\overline{\Gamma}^2$ is smaller than or of order $\overline{\Gamma}^2$ so that $\overline{\Gamma}_{\exp} \approx \overline{\Gamma}$, which is the result that we would obtain from the experiments at $\Delta\tau = 0$. However, the inequality $\Gamma_n\Delta\tau \ll 1$ does not hold for all the internal modes. Due to this the situation is quite different in the case of polymer coils. Equation (53) thus does not follow from (52). Using $\overline{\Gamma}_{\exp} = \overline{\Gamma}$, we introduce in the analysis an error, which has been estimated in the work [13] (see also [43]) as $\overline{\Gamma}_{\exp}-\overline{\Gamma} \approx (\overline{\Gamma}\Delta\tau)^{1/3}\overline{\Gamma}$. However, this estimation was done for the case of large scattering vectors; it is thus inapplicable for large-time motion and small scattering vectors, $kR \ll 1$. In the latter case the role of the internal modes is negligible and only the diffusion of the coil determines the Van Hove function. We thus argue that the reason for the observed discrepancy between the theory and experiment has a deeper origin. Probably it is due to the fact that the results describing the scattering on the coil are 1) not applicable in the limit $t \to 0$, and 2) even if the time resolution $\Delta\tau$ is relatively large, for polymer coils with large radii $R$, $\Delta\tau$ can be not enough large in the sense that the polymer motion follows the long-time regime predicted by the original RZ theory. There is a transition regime between the short-time behavior (which is very different from that in the RZ models) and the long-time behavior of the correlation functions that determine the DSF $G(k,t)$. Even for the times $t \gg \tau$ there are still long-time tails in $G(k,t)$, which make it different from the scattering function of the diffusion type, ~ exp(-$k^2Dt$). As a result, the DSF decays more slowly in the time, which could lead to an experimentally observed (apparent) diffusion coefficient that is smaller than the value $D$, predicted by the RZ theory. The situation is similar to that in the experiments on the Brownian motion of rigid particles, see, e.g., Ref. [25].

**THE FIRST CUMULANT OF THE DSF**

As discussed above, the effects of the hydrodynamic memory on the relaxation of the polymer internal modes can hardly be observed - the time dependence of the correlation functions for these modes only very slightly differs from that in the traditional RZ theory. Presumably, these effects could be revealed in the diffusion of the coil as a whole. In this case the memory effects seem to be not only measurable but could contribute to some of the existing problems between the theory and experiment.

As is well known, the diffusion of the polymer as a whole is observed at small scattering vectors, $kR \ll 1$, when the DSF can be approximated by the function

$$G(k,t) = G(k,0)\exp\left\{-k^2[\psi_0(0) - \psi_0(t)]\right\}. \tag{54}$$

The first cumulant that characterizes the decay rate of the DSF at $t \to 0$, and is defined as [1]

$$\Gamma = -G(k,0)^{-1}[dG(k,t)/dt]_{t=0}, \tag{55}$$

can be written in the form

$$\Gamma = k^2 \frac{d}{dt}[\psi_0(0) - \psi_0(t)], \quad t \to 0, \tag{56}$$

and expressed through the time-dependent diffusion coefficient $D(t) = -d\psi_0(t)/dt$ [22]. The function $\psi_0(t)$ can be found using Eq. (15) or the subsequent results for the Rouse and Zimm models. Another way is to evaluate $\Gamma$ directly from the expression for the DSF.

*The cumulant in the traditional Rouse model*

The DSF is calculated from Eq. (36). The static structure factor (SSF) can be from this equation expressed as

$$G(k,0) = \frac{1}{N}\sum_{nm}\exp\left\{-\frac{Na^2k^2}{3\pi^2}\sum_{p=1}^{\infty}\frac{1}{p^2}\left(\cos\frac{\pi np}{N} - \cos\frac{\pi mp}{N}\right)^2\right\}. \tag{57}$$

We use $Na^2 = 128R^2/(3\pi)$, and the relation between the hydrodynamic and gyration radii, $R_G = 8R/(3\sqrt{\pi})$, $R_G^2 = a^2N/6$. The sum in the exponent is $\pi^2|n-m|/(2N)$ [39]. As usually in the continuum approximation, the summation through $n$ and $m$ is replaced by the integration from 0 to $N$, so that for the SSF one obtains [1]

$$G(k,0) = \frac{2}{(kR_G)^2}\left\{1 - \frac{1}{(kR_G)^2}\left(1 - e^{-k^2 R_G^2}\right)\right\}. \tag{58}$$

Since the internal modes relax exponentially, $\psi_p(t) = \psi_p(0)\exp(-t/\tau_p)$, $p > 0$, and for the Einstein diffusion of the whole coil, the quantity $d\Phi_{nm}(t)/dt$ in Eqs. (36) and (37) is

$$\left(\frac{d\Phi_{nm}(t)}{dt}\right)_{t=0} = 2D\left[1 + 2\sum_{p=1}^{\infty} \cos\frac{\pi n p}{N}\cos\frac{\pi m p}{N}\right], \tag{59}$$

where we have used $\psi_p(0)/\tau_p = D/2$. The time derivation of the DSF can be expressed in the form

$$\left(-\frac{dG}{dt}\right)_{t=0} = \frac{k^2 D}{N}\sum_{nm}\left(\frac{d\Phi_{nm}(t)}{dt}\right)_{t=0}\exp(-k^2 R_G^2 |n-m|/N). \tag{60}$$

Calculating the sum we again convert it to the integral. The integral is easily found noting that the quantity in the square brackets in Eq. (59) is proportional to the delta function of the difference $(m - n)$,

$$N\delta(n-m) = 1 + 2\sum_{p=1}^{\infty}\cos\frac{\pi n p}{N}\cos\frac{\pi m p}{N},$$

represented through the orthonormal set of functions in the interval from 0 do $N$,

$$\psi_p(n) = (2 - \delta_{p0})^{1/2}\cos\frac{\pi n p}{N}, \qquad p = 0,1,2,...$$

Instead of Eq. (60) we thus have

$$-\frac{\partial G}{\partial t}\bigg|_{t=0} = k^2 D\int_0^N dn\int_0^N dm\,\delta(n-m)\exp\left(-\frac{k^2 R_G^2}{N}|n-m|\right) = k^2 ND \tag{61}$$

and the first cumulant from Eq. (55) is

$$\frac{\Gamma}{k^2 D} = \frac{1}{2}\frac{x^2}{x + \exp(-x) - 1}, \qquad x = (kR_G)^2. \tag{62}$$

The limiting cases of this expression are well known [1, 2]

$$\frac{\Gamma}{k^2 D} \approx 1 + \frac{x}{3} + ..., x \to 0, \tag{62a}$$

$$\frac{\Gamma}{k^2 D} \approx \frac{x}{2}, \quad x \gg 1. \tag{62b}$$

The first cumulant in the Rouse model does not reflect the internal dynamics of the polymer coil.

*The Zimm cumulant in the classical model*

We proceed analogously as in the Rouse case; the difference is only in the diffusion coefficient $D$ and in the relaxation times $\tau_p$ (20). At large $kR_G$ again Eq. (60) is used, now with

$$\left(\frac{d\Phi_{nm}(t)}{dt}\right)_{t=0} = 2D\left[1 + \frac{3}{2\sqrt{2}} \sum_{p=1}^{\infty} \frac{1}{\sqrt{p}} \cos\frac{\pi np}{N} \cos\frac{\pi mp}{N}\right]. \tag{63}$$

This expression is integrated with the factor $\exp(-k^2 R_G^2 |n-m|/N)$, due to which the main contribution is given by $n \approx m$. If the summation is replaced by integration [10], one arrives at the result

$$-\left.\frac{\partial G}{\partial t}\right|_{t=0} \approx \frac{3\sqrt{\pi}}{4} \frac{kDN}{R_G}. \tag{64}$$

Using the expression for the SSF (58) at large $kR_G$, $G(k,0) \approx 2N(kR_G)^{-2}$, the well known "$k^3$ law" is obtained for the first cumulant when the hydrodynamic interaction is taken into account:

$$\Gamma \approx \frac{k_B T}{6\pi\eta} k^3. \tag{65}$$

At small wave-vector transfers, $kR_G \ll 1$, the cumulant is, as in the Rouse model, $\Gamma = k^2 D$. The question what quantity should be considered as the diffusion coefficient of the coil is the subject of many discussions [42]. The Kirkwood relation for $D$ (Eq. 18) obtained in the preaveraging approximation from the first cumulant as the limit of $\Gamma/k^2$ at $k \to 0$ is valid for very short times and differs from the long-time diffusion coefficient. Theoretical analyses, computer simulations and scattering experiments however show that the difference between these two quantities is very small (smaller than the experimental resolution). Again, the discrepancy between the theory and measured diffusion coefficients of the polymer coil has a different origin.

*The first cumulant in the model with hydrodynamic memory equals to zero*

When the hydrodynamic memory is taken into account, the first cumulant $\Gamma$, defined by Eq. (55) identically equals to zero. Really, the time derivative of the DSF from Eq. (36) is (at $t = 0$)

$$G'(k,0) = -\frac{k^2}{N} \sum_{mn} \Phi'_{mn}(0) \exp[-k^2 \Phi_{mn}(0)]. \tag{66}$$

When we express $\Phi_{mn}(t)$ through the correlation functions of the normal modes, i.e. through the MSD of the coil ($p = 0$) and the functions $\psi_{p\geq 1}(t)$ describing the internal modes, and take into account that

$$\frac{d}{dt}[\psi_0(0)-\psi_0(t)]_{t=0} = 0 \tag{67}$$

(remind that, as distinct from the linear time dependence in the previous models, the MSD is at short times $\sim t^2$), we obtain

$$\Phi'_{mn}(0) = -4\sum_{p=1}^{\infty}\psi'_p(0)\cos\frac{\pi np}{N}\cos\frac{\pi np}{N}. \tag{68}$$

In the classical RZ theory $\psi'_p(0)$ is nonzero. Now $\psi'_p(0) = 0$, as easily seen from Eq. (15). Due to this also $\Phi'_{mn}(0)$ and $G'(k,0)$ equal to zero, consequently, $\Gamma = 0$. Illustrative numerical results for the time derivative of the correlator $\psi_0(t)$ that is connected with the time-dependent diffusion coefficient $D(t)$ are given in Fig. 2. The fact that in experiments the first cumulant is nonzero can be explained as follows. Every experiment has its "time zero". It is given by the experimental resolution in the time. The time $t = 0$, used in the theory, does not exist in experimental situations. Any experiment is able to monitor the polymer dynamics only beginning from some time $t_0$. The theoretical results are thus applicable for the times $t$ larger than $t_0$. While the time-dependent diffusion coefficient is zero for $t = 0$, at $t_0 > 0$ we have (both for the diffusion of the coil and for the internal modes) a nonzero diffusion coefficient. With the time growing our result for the DSF will approach the classical one and for enough long times the two results cannot be distinguished. As well, for poor experimental resolution, large $t_0$, our results will be very close to the classical ones. Better the experimental resolution, larger the deviation of our theory from the prediction for the first cumulant in the previous theory. Since the classical results are in our theory approached very slowly in the time, we believe that the deviations could be experimentally accessible. There are two questions to be answered. First, whether the time resolution in the current experiments is good enough to detect these deviations, and second, whether the theory is able to resolve the existing discrepancies with experiment, mentioned in Introduction. As to the first question, at least the nondiffusive motion of the whole polymer should be observable. For a typical polymer 100 nm in radius, in water solution at room conditions, the characteristic time of the ballistic motion is $\tau_R \sim 10$ ns and the transition to the diffusive regime is very slow. In the current scattering experiments the sampling times are on the level of tenths μs, but can be an order shorter [44]. Even much shorter times are accessible by the neutron spin echo technique [45]. For some of the experiments in which the ballistic motion of small rigid spheres has been successfully observed see Refs. [23 - 25]. Before answering the second question we have to calculate the DSF and the first cumulant as they follow from our hydrodynamic theory.

The relaxation of the internal modes will be assumed indistinguishable from the exponential relaxation in the previous RZ theory, $\psi_p(t) \sim \exp(-t/\tau_p)$. In this case the time derivative of the DSF (36) is

$$\frac{\partial G}{\partial t} = \frac{k^2}{N}\sum_{mn}\left\{\frac{d}{dt}\psi_0(t) - 4\sum_{p=1}^{\infty}\frac{\psi_p(0)}{\tau_p}e^{-t/\tau_p}\cos\frac{\pi np}{N}\cos\frac{\pi np}{N}\right\}\exp(-k^2 R_G^2|n-m|/N), \tag{69}$$

and for the first cumulant (55) we find at $t = t_0$

$$\frac{\Gamma}{k^2 D} = -\frac{1}{k^2 DG(k,0)} \frac{\partial G}{\partial t}\bigg|_{t=t_0} \approx -\frac{1}{D}\frac{d}{dt}\psi_0(t)\bigg|_{t=t_0} - 1 + \frac{N}{G(k,0)}. \tag{70}$$

In the Rouse limit the last term in this equation is $x^2(x-1+\exp(-x))^{-1}/2$, $x = (kR_G)^2$, and the first term in the classical theory was equal to 1. When $t_0$ increases, the contribution of the first two terms decreases. Similarly, in the Zimm limit we obtain

$$\Gamma = -k^2 D\left\{1 - D^{-1}\frac{d}{dt}[\psi_0(0) - \psi_0(t)]\bigg|_{t=t_0}\right\} + \Gamma_Z, \tag{71}$$

where $\Gamma_Z(k)$ is the cumulant in the original Zimm theory. In general, we have that in the case of "ideal" experiment, with $t_0 = 0$, the hydrodynamic memory would lower the measured cumulant by the quantity $-k^2 D$. In the case of small $kR_G$, when the pure diffusion is observed, the cumulant would be zero. Since $t_0 \neq 0$, the cumulants are nonzero but their values are smaller than in the traditional theory. The importance of this effect could be judged by a detailed comparison with experiments.

The first cumulant of long flexible polymers in $\theta$ solutions was experimentally studied in a number of works. One of the most detailed studies is the work by Sawatari *et al.* [46], which probably supports our view on the dynamics of individual polymers. While most of the investigations have been devoted to the confirmation of the universal behavior of $\Gamma$ as a function of the scattering vector $k$, the mentioned work tests the dependence of $\eta\Gamma/k_B T k^3$ on $kR_G$ in the "$k^3$ region" for different polymers. Indeed, it has been found that the cumulants differ for different polymers, even if they have large molecular weight. This is in agreement with the theoretical prediction [47], the theory is however not able to explain quantitatively the data for an individual polymer.

It is not very surprising that the "universal" region is in fact not universal (i.e. that the Zimm plot $\eta\Gamma/k_B T k^3$ on $kR_G$ depends on the system polymer – solvent). Such universality requires the existence of a "pure" Zimm polymer while the dynamics of every polymer, within the standard bead-spring model, reveals at the same time properties of both the Zimm and Rouse polymers. Different polymers are thus described (in addition to the parameters entering the Zimm model) by different phenomenological friction coefficients for one bead that could be the reason for the observed nonuniversality [34]. The importance of the work [46] is also in the detailed investigation of the following two problems in the experimental determination of the first cumulant. First, it is known that the determination of $\Gamma(k)$ is sensitive to the experimental sampling time [12]. It seems also that it depends on the choice of the method of its evaluation from the data [46]. For us both these problems are important since the effect of hydrodynamic memory is detectable only at short times, on the level of the shortest resolution times in the usual light scattering experiments. In the Sawatari's experiments solutions of polystyrene and poly(methylmethacrylate) of large molecular weights were studied by the static and dynamic light scattering. The diffusion coefficients of the observed polymer coils have been compared to their gyration radii. The values of $D$ were always smaller than the corresponding Kirkwood values for the Zimm model. The DSF was determined from the normalized autocorrelation function of the scattered light, $g^{(2)}(t)$, as $G(k,t) \propto g^{(2)}(t) - 1$. In our approach the measured function $g^{(2)}(t)$ corresponds to

$$\ln[g^{(2)}(t)-1] = \text{const} + k^2\{D_C t - [\psi_0(0) - \psi_0(t)]\} + \ln G(k,t), \tag{72}$$

if $G$ is understood as the DSF in the previous RZ theory. The additional second term on the right, which disappears at long times, determines the deviation from the value expected within the RZ theory. The differences can be detected only at short sampling times. Experimental results are in qualitative agreement with our predictions. Really, the measured values of $\ln[g^{(2)}(t) - 1]$ obtained with the sampling time $t_0 = 0.5$ μs were slightly larger than the DSF values in the region of its decay at short times, found with a longer $t_0 = 2$ μs. The reason for this difference was unclear for the authors [46] (note that for a different polymer, using closer sampling times, the difference in the values of $g^{(2)}(t)$ was smaller). The observed difference is small (since the used times $t_0$ are still rather long compared to the characteristic time $\tau_R$, which we have assumed for the longest polymer to be $\tau_R \sim 10^{-8}$ s). Although the quantitative comparison with the experiment requires a precise treatment of the experimental data using the above derived expression for the DSF, the noted observations support our theory. As to the first cumulant $\Gamma$ determined from $g^{(2)}(t)$ by the extrapolation of $(1/2)\ln[g^{(2)}(t) - 1]$ to $t = 0$, only the data for $t > 5$ μs have been considered in the determination of $\Gamma$, and the diffusion coefficient was measured with the sampling time 22–27 μs, i.e. far out of the region of our main interest. The observed deviations from the values $\eta\Gamma/k_B T k^3 = 1/6\pi$ and $D$ expected in the frame of the Zimm theory were small, of order of 10%. The memory effects do not persist at such long times (for the considered situation our theory predicts corrections to the RZ results for $\Gamma$ and $D$ on the level of 2% and 1%, respectively), but another reasons can play a role (e.g., the mentioned combined Rouse-Zimm behavior of the polymers, or the approximate character of the used expressions in the region of relatively small ($2 < kR_G < 7$) scattering wave vectors). It thus seems that the hydrodynamic memory alone cannot give a satisfactory solution of the long-standing puzzles discussed in Introduction. We however believe that the presented hydrodynamic approach could stimulate new studies on the dynamics of polymers in solution, similarly as it became more than three decades ago in the theory of the Brownian motion.

**EFFECTS OF HYDRODYNAMIC NOISE ON THE POLYMER DYNAMICS**

In the preceding consideration the random forces acting on the polymer elements were given by Eq. (14). This is not the only way how to determine the spectral properties of these forces in the theory of the Brownian motion. In what follows we shall study the properties of the random forces based on a different approach. At the motion of the spherical particle in a liquid we shall take into account, along with the velocity field created by the motion of this particle, the additional velocity and pressure fields due to the spontaneous fluctuations of the tensor of random stresses, $S_{\alpha\beta}$ (the spontaneous hydrodynamic noise). The noise is assumed to be Gaussian, with the first moment equal to zero, and the quadratic fluctuations of the tensor will be traditionally [48] defined by the delta-correlated expression

$$\langle S_{\alpha\beta}(\vec{r},t) S_{\alpha'\beta'}(\vec{r}',t') \rangle = 2k_B T \eta \left( \delta_{\alpha\alpha'} \delta_{\beta\beta'} + \delta_{\alpha\beta'} \delta_{\alpha'\beta} - \tfrac{2}{3} \delta_{\alpha\beta} \delta_{\alpha'\beta'} \right) \delta(\vec{r} - \vec{r}')\, \delta(t - t'). \tag{73}$$

Let the velocity $\vec{v}^{\omega}(\vec{r})$ and the pressure $p^{\omega}(\vec{r})$ be the Fourier components of the field of hydrodynamic noise, which is created by the random stresses $S_{\alpha\beta}^{\omega}$ in the absence of the particle, and the fields in the presence of the particle moving with the velocity $\dot{\vec{x}}_n$ will be denoted by $\vec{V}^{\omega}(\vec{r})$ and $P^{\omega}(\vec{r})$. The origin of the spherical system of coordinates will be chosen in the center of inertia of the particle. The boundary problem for the determination of the Fourier components of the velocity and pressure are written in the following form:

$$-i\omega\rho\vec{V}^{\omega} = -\nabla P^{\omega} + \eta\Delta\vec{V}^{\omega} + \vec{F}^{\omega}, \quad F_{\alpha}^{\omega} = \nabla_{\beta}S_{\alpha\beta}^{\omega}, \quad \mathrm{div}\vec{V}^{\omega} = 0 \qquad (74)$$

$$\vec{V}^{\omega}(r=b) = \dot{\vec{x}}_n^{\omega}, (|\vec{r}-\vec{x}_n|=b); \quad \vec{V}^{\omega}(\vec{r}) \to \vec{v}^{\omega}(\vec{r}), \ (r \gg b). \qquad (75)$$

The solution of a similar problem is given in the works by Bedeaux and Mazur [49, 50], where it was used for the determination of the tensor of stresses and the hydrodynamic force acting on the particle. We use these results and represent the hydrodynamic force on the elements of the polymer chain in the form of two contributions. The first of them coincides with the nonstationary expression (2), and the second one is expressed as a random force, the properties of which are determined by the correlators (73),

$$\vec{f}_n^{\omega} = \xi\left[(1+b\chi)\,\vec{v}^{S\omega}(\vec{x}_n) + \tfrac{1}{3}b^2\chi^2\,\vec{v}^{V\omega}(\vec{x}_n)\right] \qquad (76)$$

($\chi$ is defined after Eq. (7)). Here the following integrals over the surface and volume of the particle are introduced, the center of inertia of the particle being placed in the point $\vec{x}_n$:

$$\vec{v}^{S\omega}(\vec{x}_n) = S^{-1}\int \vec{v}^{\omega}(\vec{x}_n + b\vec{n}_0)dS, \quad \vec{v}^{V\omega}(\vec{x}_n) = V^{-1}\int \vec{v}^{\omega}(\vec{x}_n + \vec{r})dV. \qquad (77)$$

Note that integrating at $\vec{x}_n = 0$ for the bilinear averages of the random force (76) and using (73), the result [49] coincides with the traditional one, based on the FDT [22]

$$\langle f_{\alpha}^{\omega} f_{\beta}^{\omega'} \rangle = 2k_B T\,\mathrm{Re}\,\xi^{\omega}\delta_{\alpha\beta}\delta(\omega+\omega'). \qquad (78)$$

We use the results (2), (76), and (77) for the construction of bilinear spectra of the internal amplitudes of the polymer chain,

$$\langle y_{p\alpha}^{\omega} y_{q\beta}^{\omega'} \rangle = \frac{\langle f_{p\alpha}^{\omega} f_{q\beta}^{\omega'} \rangle}{\left(-i\omega\Xi_p^{\omega} - M\omega^2 + K_p\right)\left(-i\omega'\Xi_q^{\omega'} - M\omega'^2 + K_q\right)} \qquad (79)$$

(the quantities $\Xi$ and $K_q$ are defined in Eq. (12), and $M$ is the mass of the bead), which are connected to the spectral densities of the noise by the integral transformation

$$\langle f_{p\alpha}^{\omega} f_{q\beta}^{\omega'} \rangle = \frac{1}{N^2}\int_0^N dn\int_0^N dm\,\langle f_{n\alpha}^{\omega} f_{m\beta}^{\omega'} \rangle\cos\frac{\pi p n}{N}\cos\frac{\pi q m}{N}. \qquad (80)$$

The quadratic fluctuations (73) of the stress tensor are delta-correlated so that we can write

$$\langle f^{\omega}_{n\alpha} f^{\omega'}_{m\beta} \rangle = \delta(\omega+\omega') \langle \widehat{A}\widehat{A}' v^{\omega}_{\alpha}(\vec{x}_n + \vec{r}) v^{\omega'}_{\beta}(\vec{x}_m + \vec{r}') \rangle, \quad \vec{f}^{\omega}_n = \widehat{A}\vec{v}^{\omega}(\vec{x}_n + \vec{r}), \tag{81}$$

where we have introduced the operator $\widehat{A}$ acting according to the rule (76). The spectral density of the fluctuations of the velocity field due to the influence of noise is determined by integrating the hydrodynamic susceptibility [48]

$$\langle v^{\omega}_{\alpha}(\vec{R}) v^{\omega*}_{\beta}(\vec{R}') \rangle = \delta_{\alpha\beta} \frac{k_B T}{12\pi^3 \rho} \int \frac{\nu k^2 \exp(i\vec{k}(\vec{R}-\vec{R}'))}{\omega^2 + \nu^2 k^4} d\vec{k}, \tag{82}$$

where $\nu$ is the kinematic viscosity. Before performing the double integration in Eq. (80) over the discrete variables (continuum approximation), we average the exponential in Eq. (82) over the equilibrium distribution function $P(r_{nm})$ (see Eq. (9)) of the chain elements (as in the case of the Oseen tensor):

$$\langle \exp(i\vec{k}(\vec{x}_n - \vec{x}_m)) \rangle_0 = \int \exp(i\vec{k}\vec{r}_{nm}) P(r_{nm}) d\vec{r}_{nm} = \exp\left[-\frac{k^2 a^2}{6}|n-m|\right]. \tag{83}$$

From here, after the integral transformation we have in the same approximation as for the Oseen matrix elements $h_{pq}$ (Eqs. (13), (27))

$$\left[\langle \exp(i\vec{k}(\vec{x}_n - \vec{x}_m)) \rangle_0\right]_{pq} \approx \delta_{pq} \frac{24}{pNa^2} \frac{k^2}{k^4 + (6\pi p/Na^2)^2}. \tag{84}$$

Now Eq. (80) can be written in the form

$$\langle f^{\omega}_{p\alpha} f^{\omega'}_{q\beta} \rangle \approx \delta(\omega+\omega')\delta_{pq}\delta_{\alpha\beta} \frac{2k_B T}{\pi^3 pNa^2\eta} \int d\vec{k} \frac{\widehat{A}\widehat{A}'^* \exp(i\vec{k}(\vec{r}-\vec{r}'))}{k^4 + (\omega/\nu)^2} \frac{k^4}{k^4 + (6\pi p/Na^2)^2}. \tag{85}$$

After the action of the operators $\widehat{A}$ we have

$$\widehat{A}\exp(i\vec{k}\vec{r}) = \zeta\left[(1+b\chi)\frac{\sin bk}{bk} + (b\chi)^2\left(\frac{\sin bk}{bk} - \cos bk\right)(bk)^{-2}\right]. \tag{86}$$

Now the spectral density of the amplitudes $y^{\omega}_p$ can be written for $p = 1, 2, \ldots$ as

$$\langle |y^{\omega}_p|^2 \rangle = \frac{24 k_B T b}{\pi^2 pNa^2\eta} \int_0^{\infty} dk \frac{k^6}{k^4 + (\omega b^2/\nu)^2} \frac{1}{k^4 + (6\pi p b^2/Na^2)^2}$$

$$\times \left|(1+b\chi)\frac{\sin k}{k} + (b\chi)^2\left(\frac{\sin k}{k^3} - \frac{\cos k}{k^2}\right)\right|^2 \left|\frac{\xi}{-i\omega \Xi^{\omega}_p - M\omega^2 + K_p}\right|^2. \tag{87}$$

The integration over the dimensionless variable can be performed using the decomposition of the integrand in simple fractions. The answer can be expressed in terms of elementary functions and the error function. It is rather cumbersome so we shall not give it here. Instead we shall consider some more simple consequences of the theory.

To investigate the diffusion motion of the polymer coil as a whole one should analyze the dynamical properties of the radius vector of the center of inertia,

$$\vec{y}_0^\omega = \tfrac{1}{N}\int_0^N dn\, \vec{x}^\omega(n).$$

The diffusion coefficient of the coil can be found using the Kubo formula

$$D = \frac{1}{3}\int_0^\infty \langle \dot{\vec{y}}_0(t)\dot{\vec{y}}_0(0)\rangle dt, \tag{88}$$

Equivalently, it can be expressed through the spectral density $\langle |\dot{\vec{y}}_0^\omega|^2 \rangle|_{\omega=0}$. Using Eqs. (11), (12), and (76), we obtain

$$\langle |\dot{\vec{y}}_0^\omega|^2 \rangle = \frac{k_B T}{\pi^2 N^2 \rho}\int_0^N dn \int_0^N dm \int_0^\infty dk\, \frac{\nu k^4}{\omega^2 + (\nu k^2)^2}\left(\frac{\sin kb}{kb}\right)^2 \exp(-k^2 a^2 |n-m|/6). \tag{89}$$

The Zimm diffusion coefficient can be easily obtained from the Kubo relation at $b = 0$:

$$D_Z = \frac{k_B T}{3\pi^2 N^2 \eta}\int_0^N dn \int_0^N dm \int_0^\infty dk\, \exp(-k^2 a^2 |n-m|/6) = \frac{k_B T}{\sqrt{6\pi^3 N^2 \eta a}}\int_0^N dn \int_0^N \frac{dm}{\sqrt{|n-m|}}. \tag{90}$$

The double integral equals to $8N^{3/2}/3$ and we get the correct diffusion coefficient (18) in the Zimm limit. At $b > 0$ we introduce the dimensionless parameter $\sigma^2 = 6b^2/(a^2 N)$ and use

$$\int_0^N dn \int_0^N dm\, f(|n-m|) = \int_0^N dn \left\{\int_0^n ds\, f(s) + \int_0^{N-n} ds\, f(s)\right\} = 2\int_0^N (N-s)f(s)ds.$$

For $f(s) = \exp(-k^2 s/\sigma^2 N)$ the double integral from 0 to $N$ in Eq. (89) will be

$$\frac{2\sigma^2 N^2}{k^2}\left(1 - \frac{\sigma^2}{k^2} + \frac{\sigma^2}{k^2}\exp(-k^2/\sigma^2)\right).$$

Then the diffusion coefficient is expressed as

$$D = D_Z \Psi(\sigma), \qquad \Psi(\sigma) = \frac{3}{2\sqrt{\pi}}\int_0^\infty \frac{dx}{x^4}\left(\frac{\sin\sigma x}{\sigma x}\right)^2 (x^2 - 1 + e^{-x^2}), \tag{91}$$

with the limiting value $\Psi(0)=1$ at $\sigma = 0$, when $D = D_Z$. Integrating Eq. (91) *per partes*, the function $\Psi$ can be expressed in terms of elementary functions and the error function. The effective hydrodynamic radius of the coil, $R_C \sim a\sqrt{N}/\psi(\sigma)$, obtained taking into account the fluctuations of the random stress tensor, contains a weak dependence on the ratio between the size of the bead $b$ and the size of the coil $a\sqrt{N}$.

Let us also consider the velocity autocorrelation function of the coil,

$$\Phi_0(t) = \langle \dot{\vec{y}}_0(t)\dot{\vec{y}}_0(0)\rangle = \frac{1}{2\pi}\int d\omega \cos\omega t \langle |\dot{\vec{y}}_0^\omega|^2\rangle. \tag{92}$$

At long times it is sufficient to restrict ourselves to small $\omega$. Again, the most simple case corresponds to the Zimm limit $b = 0$. Equations (79) - (86), using the found double integral over $n$ and $m$ and the integration over $\omega$ yield

$$\Phi_0(t) = \frac{6k_B T}{\pi^2 \rho N a^2}\int_0^\infty dk \exp(-k^2 \nu t)\left(1 - \frac{6}{Na^2 k^2} + \frac{6}{Na^2 k^2}\exp(-k^2 Na^2/6)\right). \tag{93}$$

The result of integration is

$$\Phi_0(t) = \frac{3k_B T}{\pi^{3/2}\rho Na^2}\left\{\frac{1}{\sqrt{\nu t}} + \frac{12}{Na^2}\left[\sqrt{\nu t} - \sqrt{\nu t + Na^2/6}\right]\right\}. \tag{94}$$

It gives the correct asymptote at long times $t \gg Na^2/6\nu$, which is independent on the polymer parameters,

$$\Phi_0(t) \approx \frac{k_B T}{4\rho(\pi\nu t)^{3/2}}. \tag{95}$$

This expression is the same as for an individual Brownian particle when the viscous aftereffect is taken into account [51].

At shorter times $\nu t \ll Na^2/6$ (assuming however $t \gg b^2\rho/\eta$) we find for the Zimm polymer $\Phi(t) \sim t^{-1/2}$ [52]

$$\Phi_0(t) \approx \frac{3k_B T}{Na^2}\frac{1}{\sqrt{\pi^3 \rho \eta t}}. \tag{96}$$

For arbitrary $b$ the VAF is expressed through the integral

$$\Phi_0(t) \propto \int_0^\infty dk \left(\frac{\sin kb}{kb}\right)^2 \exp(-k^2 \nu t)\left(1 - \frac{6}{Na^2 k^2} + \frac{6}{Na^2 k^2}\exp(-k^2 Na^2/6)\right). \tag{97}$$

The integral can be evaluated and expressed in terms of the error function and elementary functions. For example, introducing

$$x_1 = \frac{b}{\sqrt{vt}}, \quad x_2 = \frac{b}{\sqrt{vt + Na^2/6}}, \quad \varphi(x) = \mathrm{erf}(x) - \frac{1}{\sqrt{\pi}x}\left(1 - e^{-x^2}\right),$$

the VAF can be written in the form

$$\Phi_0(t) = \frac{6k_B Tb}{\pi \rho N^2 a^4}\left\{\frac{Na^2}{2b^2}\varphi(x_1) + \frac{1}{x_1^2}\mathrm{erf}(x_1) - \frac{1}{x_2^2}\mathrm{erf}(x_2)\right.$$

$$\left. + 2\left[\left(1 + \frac{1}{x_1^2}\right)\varphi(x_1) - \left(1 + \frac{1}{x_2^2}\right)\varphi(x_2)\right] + \frac{1}{\sqrt{\pi}}\left(\frac{1}{x_1} - \frac{1}{x_2}\right)\right\}. \tag{98}$$

At $b \to 0$ we return to the previous result (94). At $t \to 0$ we have from Eq. (97)

$$\Phi_0(0) = \frac{3k_B T}{M_{\mathrm{eff}}}, \tag{99}$$

where the "effective" mass of the polymer coil is given by

$$\frac{1}{M_{\mathrm{eff}}} = \frac{1}{\pi \rho Na^2 b}\left\{\left(1 + \frac{2}{3}\sigma^2\right)(1 - \mathrm{erf}\,\sigma) - \frac{2\sigma}{3\sqrt{\pi}}\left[1 + \frac{1 + \sigma^2}{\sigma^2}\left(e^{-\sigma^2} - 1\right)\right]\right\}. \tag{100}$$

The Rouse limit $\sigma^2 \gg 1$ of this equation is

$$M_{\mathrm{eff}} = \frac{\pi \rho N^2 a^4}{4b}.$$

### SINGLE POLYMER DYNAMICS IN THE PRESENCE OF OTHER POLYMERS

The theory presented in the previous paragraphs is developed for a single polymer chain or very dilute solutions and does not contain any dependence on other polymers in solution. Our aim of this paragraph is to show that the above results can be relatively simply generalized to take into account the presence of other polymer coils through their finite concentration in solution. We shall focus on dilute polymer solutions where the coils are well separated. That is, we do not consider semidilute and dense solutions with the chains strongly overlapping each other, so that the polymers lose their individualities. The static properties of dilute polymer solutions are thought to be well understood [1, 2]. However, still problems exist in understanding, e.g., the viscosity behavior of such solutions [53]. In what follows we shall give an outline of a simple phenomenological theory of the diffusion of the polymer as a whole and the relaxation of its internal modes in the case when other chains in the solution affect the flow of the solvent. The results are interesting also because they imply the screening of the hydrodynamic interaction and the transition between the Zimm and Rouse dynamics of the "test" polymer. These effects are well revealed in semidilute and dense polymer solutions (see [1, 11, 54] and references there) but qualitatively they are displayed already in the theories for dilute solutions [1].

In what follows we will not take into account the effects of solvent and polymer beads inertia and consider solely the concentration dependence of the diffusion coefficients and relaxation times of the polymer internal modes. This means that we are interested in the time scale $t \gg \tau_R$, which corresponds to standard experiments. In this case we can consider the stationary hydrodynamic equations and the steady-state motion of the polymer. The theory could be generalized to the nonstationary case as well, which would allow to study the time dependence of the mentioned transition between the various regimes of the polymer dynamics. This will be the subject of our next investigation. In the present approach the distribution of the polymer coils in solution is assumed to be stationary. For the simplicity only $\theta$ solvents are considered with no excluded volume effects. The main effects will be well seen and an extension of the theory to other situations requires only replacing the used equilibrium distribution of the polymer beads with a model distribution for nonideal polymers.

The basic equations of our model are given in the first paragraph after Introduction. The equation of motion of the test polymer is Eq. (1) with the zero mass of the bead. The resistance force on the bead during its motion is given by Eq. (2). Instead of the Navier-Stokes equation (3) we shall use the equation (see the classical work by Debye and Bueche [55] and the references there to the earlier Brinkman's works)

$$\rho \frac{\partial \vec{v}}{\partial t} = -\nabla p + \eta \Delta \vec{v} - \kappa^2 \eta \vec{v} + \vec{\varphi}. \tag{101}$$

In the cited works a phenomenological model for a polymer has been introduced considering it as a porous permeable medium. In the right hand side of the Navier-Stokes equation (4) a force $-\kappa^2 \eta \vec{v}$ has been added, where $1/\kappa^2$ is called the permeability of the solvent. This force has a sense of the average value of the force acting on the liquid in an volume element $dV$, provided the average number of polymers in solution per $dV$ is $c$; then $\kappa^2 \eta = cf$, where $f$ is the friction factor on one polymer coil. We have already used such approach in modeling the dynamical properties of proteins [56, 57]. As a possible model for the polymer dynamics the Brinkman's phenomenology was discussed by the authors of Ref. [54] as a possible way to describe their computer simulations. Recently, it has been applied to the description of the dynamics of polymer in semidilute and dense solutions [58]. The application of the Brinkman model to the latter situation could be questioned. However, in the dilute regime, when the polymer chains conserve their individuality and the time scale of the motion of the whole coils and the characteristic times of the internal motion of the polymers are well separated (the relaxation times of the internal modes, $\tau_p$, are always much shorter than the diffusion time of the coil, $\tau_D$), the model of the effective permeable medium seems to be a good phenomenology. It takes into account the presence of the obstacles (other chains in solution) to the flow as distinct to the flow when only one polymer (our test polymer) is present in the solvent. As a result the flow is disturbed by other coils and at their sufficiently large concentrations it becomes "frozen", which imitate the hydrodynamic screening.

All the necessary equations are already in our disposition. Now we have to solve Eq. (101) instead of the stationary variant of Eq. (3). Using again the Fourier transformation in the time and coordinates (6), the velocity field of the solvent is expressed in the form (6) through the Oseen tensor (7). The only difference is that in the variable $y = r\chi$ the quantity $\chi$ is given by $\chi^2 = \kappa^2 - i\omega\rho/\eta$ (in the nonstationary variant of the theory) and $\chi = \kappa$ in the stationary theory considered below. The preaveraged Oseen tensor can be given the form

$$\left\langle H_{\alpha\beta}^{\omega}\right\rangle_{0}=\frac{\delta_{\alpha\beta}}{6\pi\eta}\left\langle\frac{e^{-\chi r}}{r}\right\rangle_{0}, \qquad (102)$$

showing that the quantity $1/\kappa$ (in the stationary case) can be approximately considered as a screening length. The Oseen matrix is determined by Eqs. (9) and (13) and its elements are expressed by Eqs. (9) and (26).

Let us discuss the stationary variant of the theory. In this case we can formally put $M = 0$ in the equations of motion. Moreover, the Oseen tensor will not depend on $\omega$ at all.

*Diffusion of the coil*

The diffusion of the coil is described by the same equations (17) - (20) as in the case of an individual polymer. That is,

$$\psi_0(0)-\psi_0(t)=Dt=k_BT\left(h_{00}^0+\frac{1}{N\xi}\right)t. \qquad (103)$$

Instead of Eq. (26) we have for $h_{00}^0$ at $\omega = 0$

$$h_{00}^0=\frac{2}{\sqrt{6N\pi\eta}a}\frac{1}{\chi_0}\left[1-\frac{2}{\sqrt{\pi}\chi_0}-\frac{1}{\chi_0^2}\left(\exp\chi_0^2\,\mathrm{erfc}\chi_0-1\right)\right], \qquad \chi_0=\kappa R_G. \qquad (104)$$

The diffusion coefficient now depends on the permeability (or the number of the coils per unit volume, $c$):

$$D(c)=k_BT\left(h_{00}^0+\frac{1}{N\xi}\right)=D_Z(c)+D_R \qquad (105)$$

and consists of the Rouse (independent on the presence of other polymers) term (18) and the concentration dependent Zimm contribution. With the increase of the concentration the Zimm term decreases and for very large $c$ (small permeability $\kappa$ when $\chi_0 \gg 1$) it becomes

$$D_Z(c)\approx\frac{2k_BT}{\pi\eta Na^2}\frac{1}{\kappa}. \qquad (106)$$

At zero concentrations $D_Z(0)$ is given by Eq. (18). The realistic case of small $c$ is described by the expansion of Eq. (105) for $\chi_0 = \kappa R_G \ll 1$

$$D_Z(c)=k_BTh_{00}^0(c)=D_Z(0)\left(1-\frac{3}{8\sqrt{\pi}}\kappa R_G+...\right). \qquad (107)$$

The dependence of the permeability on the concentration can be estimated as follows. The quantity $\kappa$ in Eq. (101) is given by $\kappa^2 = cf/\eta$. The friction coefficient on one coil can be determined from the Einstein relation $D = k_B T / f$, where $D = D_R + D_Z(0)$. In such a picture

$$\kappa^2 = c \left[ \frac{8}{3(6\pi^3 N)^{1/2} a} + \frac{1}{6\pi N b} \right]^{-1}. \tag{108}$$

Thus, a tendency of the transition between the Zimm and Rouse behavior of our test polymer is seen. The possibility of such a transition depends on the ratio $D_Z(0)/D_R$, which is proportional to the draining parameter introduced after Eq. (20). If this parameter is large, the diffusion of the polymer at zero concentration is of the Zimm type and, theoretically, with $c$ growing the polymer could change its behavior to the diffusion with the Rouse diffusion coefficient $D_R$. The crossover concentration can be estimated from the relation $h_{00}^0(c_{\text{cross}}) \approx 1/N\xi$, i.e.

$$\frac{1}{\chi_0} \left[ 1 - \frac{2}{\sqrt{\pi}\chi_0} - \frac{1}{\chi_0^2} \left( \exp \chi_0^2 \, \text{erfc} \chi_0 - 1 \right) \right] \approx \frac{4}{3\sqrt{\pi}} \frac{D_R}{D_Z(0)}. \tag{109}$$

When the draining parameter is small, the polymer will be for all concentrations of the Rouse type and no crossover can be observed. It would be interesting to investigate also the time dependence of the hydrodynamic screening. For the diffusion of the test polymer we cannot do it in our model since the used concentration is time-independent.

*Dynamics of internal modes*

Equation (27) for the diagonal elements of the preaveraged Oseen matrix now has the form

$$h_{pp}^0 = \frac{1}{\pi \eta a \sqrt{3\pi N p}} \frac{1 + \chi_p}{1 + (1 + \chi_p)^2}, \quad \chi_p = \sqrt{\frac{N}{3\pi p}} \kappa a, \quad p = 1, 2, \ldots \tag{110}$$

It is seen from this equation that in the stationary case ($\omega = 0$) and at $\kappa = 0$ we return to the known result [1, 2]

$$h_{pp}^0(0) = \frac{1}{(12\pi^3 N p)^{1/2} \eta a}. \tag{111}$$

Now $h_{pp}^0(c)$ depends on the concentration. The internal modes relax exponentially as given by Eq. (19), but their relaxation rates again consist of the Rouse contribution (independent on the concentration) and the concentration dependent Zimm part,

$$\frac{1}{\tau_p(c)} = \frac{1}{\tau_{pR}} + \frac{1}{\tau_{pZ}(c)}, \tag{112}$$

where $\tau_{pR}$ and $\tau_{pZ}(0)$ are given by Eq. (20) and

$$\tau_{pZ}(c) = \frac{1}{2}\tau_{pZ}(0)\frac{1+(1+\chi_p)^2}{1+\chi_p}, \qquad (113)$$

which behaves as

$$\tau_{pZ}(c) = \tau_{pZ}(0)\left(1+\frac{N}{6\pi p}\kappa^2 a^2 - ...\right) \text{ as } c \to 0, \qquad (114)$$

and (although unphysical), as $c \to \infty$ one has

$$\tau_{pZ}(c) \approx \frac{1}{2}\tau_{pZ}(0)\chi_p = \frac{(Na^2)^2 \eta}{6\pi k_B T p^2}\kappa. \qquad (115)$$

In the case of the internal modes the draining parameter is determined by the relation between the Zimm and Rouse relaxation rates, i.e. $h(p) = \tau_{pR}/\tau_{pZ}$ (for small $h$ we have the Rouse behavior). If the polymer at zero concentration behaves as the Rouse one ($\tau_p(0) \approx \tau_{pR}(0)$) then, when the concentration grows, it remains the Rouse polymer. If at $c = 0$ the polymer behavior is the Zimm-like, with the increase of the concentration a tendency to the Zimm-Rouse crossover appears. This crossover can theoretically realize approximately at $\tau_{pZ}(c) \approx \tau_{pR}$ when the Rouse contribution begins to dominate. The crossover concentration $c_{\text{cross}}$ (depending on the mode number $p$) can be then determined from Eq. (113).

In case of the internal modes also the question of the time dependence of the crossover between the Rouse and Zimm behavior of the polymer can be discussed. Consider for definiteness the MSD of the $n$th bead from $N$ beads in the chain. It has the form (see Eqs. (35) and (37))

$$\langle r_n^2(t)\rangle = 3\langle [x_n(0) - x_n(t)]^2\rangle = 6\sum_{p=0}^{\infty}[\psi_p(0) - \psi_p(t)](4 - 3\delta_{p0})\cos^2\frac{\pi n p}{N}, \qquad (116)$$

where in the long-time approximation the functions $\psi_p$ are given by Eqs. (17) and (19). In the simplest special case of the end monomer (which corresponds to the experimental situation [32]) we have

$$\langle r^2(t)\rangle = 6D(c)t + \frac{4Na^2}{\pi^2}\sum_{p=1}^{\infty}\frac{1}{p^2}\left[1 - \exp\left(-\frac{t}{\tau_p(c)}\right)\right], \qquad (117)$$

with the diffusion coefficient and relaxation rates from Eq. (105) and (112). Since the relaxation rates increase with $p$, beginning from some $p$ all the modes will be of the Rouse type, having small draining parameter $h(p)$. It can be seen from Eq. (117) that at short times the contribution to the MSD determined by the internal modes is dominated by the Rouse modes with $\tau_{pR} \sim p^{-2}$. At sufficiently long times the Zimm modes give the main contribution to the MSD. It is demonstrated by Fig. 6 showing the time-dependent contribution of the Rouse modes relative to the total MSD due to the internal modes, $F_R(t)$. The calculations were done according to the formulas

$$F_R(t) = \left[ f_\infty(t) - f_{h^2}(t) \right] / f_\infty(t), \qquad f_{h^2}(t) = \sum_{p=1}^{h^2} p^{-2} \left[ 1 - \exp(-t/\tau_p) \right]. \tag{118}$$

and the modes with $h(p) < 1$ have been considered as the Rouse modes.

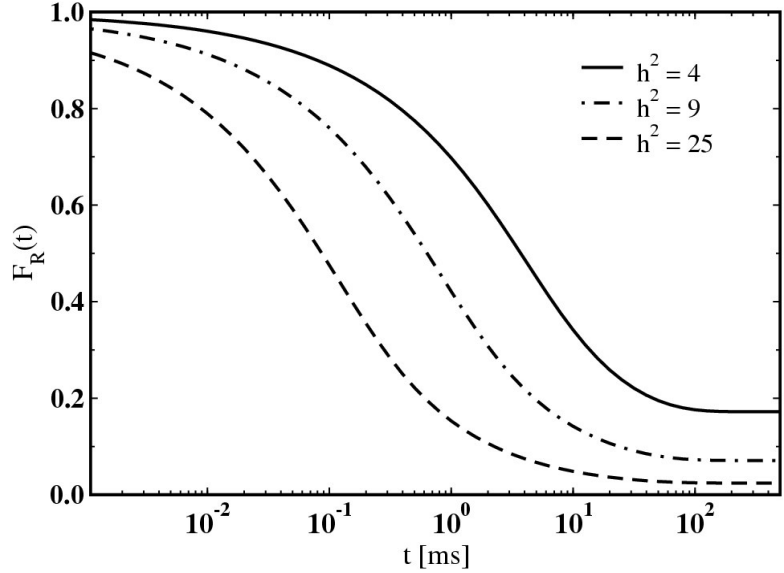

**Fig. 6** Time dependence of the contribution of the internal "Rouse" modes with $p > h^2$ ($h$ is the draining parameter for the lowest Rouse mode) relative to the total MSD due to the internal motion, calculated from Eq. (118). The polymer parameters are the same as in Fig. 1.

Thus the MSD changes its behavior in the time from the Rouse to Zimm type dynamics. When the concentration of the polymers in the solution increases, the type of the dynamics changes in the opposite way. Consequently, the observed dynamics in the solutions with nonzero concentration is determined by interplay of these two processes. At short times we have always the Rouse-type internal dynamics of the polymer, even if the draining parameter $h(p = 1)$ indicates the Zimm type dynamics. At long times every polymer behaves as the Zimm one (as far as we are concerned with the internal modes). When the concentration increases sufficiently, the dynamics has a tendency to change to the Rouse dynamics. The Rouse-like behavior has been confirmed experimentally [1] through the observed dependence $\tau_p \sim p^{-2}$. As distinct from the effective medium theory [1], in our theory the relaxation times for infinite concentrations become *exactly* the Rouse times, $\tau_p = \tau_{pR}$. The presented theory could be used to interpret computer simulations studies similar to those on semidilute polymer solutions [54]. The concentration effects should be also taken into account in the interpretation of real experiments, such as the dynamic light and neutron scattering. The corresponding formulas are given in the paragraph devoted to the DSF, with the simple replacement of the diffusion coefficient and relaxation times by the concentration-dependent quantities (105) and (112).

**CONCLUSION**

In conclusion, the main result of this work is a generalization of the popular Rouse-Zimm theory of the dynamics of flexible polymers in solution by taking into account the hydrodynamic memory (the viscous aftereffect), which is a consequence of fluid inertia. This has led to several interesting peculiarities in the time correlation functions describing the polymer motion. When the memory of the viscous solvent is taken into account, the time behavior of these functions essentially differs from that in the original theory. The mean square displacement o the whole coil is at short times proportional to $t^2$, instead of $\sim t$. At long times it contains additional (to the Einstein term) contributions, the leading of which is $\sim \sqrt{t}$. The internal normal modes of the polymer motion now do not relax exponentially. It is displayed in the long-time tails of their time correlation functions. The longest-lived contribution to the correlation function of the bead displacement is $\sim t^{-3/2}$ in the Rouse limit and $t^{-5/2}$ in the Zimm case, when the hydrodynamic interaction is strong. It would be interesting to investigate the found peculiarities using computer simulation methods and experimentally, e.g., by the dynamic light or neutron scattering. While in the case of the internal modes the differences from the original theory could hardly be observed, the "ballistic" motion of the center of inertia of the polymer should be experimentally accessible. Note that the algebraic tail $\sim t^{-3/2}$ that we have found in the time decay of the autocorrelation function of the polymer velocity [26 - 29] has been recently confirmed in computer simulations of the dynamics of individual polymer chains in solution [59].

In order to have a possibility to compare the theory with experiments, we have calculated the dynamic structure factor (DSF) of the polymer coil in various scattering regimes. We have determined the corresponding first cumulants for the Rouse and Zimm polymers. The relation between our theory and experiments is discussed in detail. The measured values of the diffusion coefficients and the first cumulants to the polymer DSF are smaller than it has been predicted by the previous theory. We have shown that our results are (at least qualitatively) in agreement with the experimental observations. The importance of our results for the description of the experiments can be however judged only after a detailed analysis of the experimental data. We discuss that such a comparison should come from the joint Rouse-Zimm model, instead of its limiting cases as it is usually done.

We have also derived the generalized Rouse-Zimm equation by a different way, taking into account the effects of hydrodynamic noise. As the random forces responsible for the noise, the fluctuations of the hydrodynamic stress tensor are taken. As a result, the spectral properties of the random forces acting on the polymer segments are not delta-correlated and are determined by the hydrodynamic susceptibility of the solvent. The preaveraging of the Oseen tensor for the nonstationary Navier-Stokes equation allowed us to relate the time correlation functions of the Fourier components of the segment radius vector to the correlation functions of the hydrodynamic field created by the noise. The velocity correlation function of the center of mass of the coil has been considered in detail. At long times its behavior follows the algebraic $t^{-3/2}$ law and does not depend on any polymer parameters.

Finally, the influence of other coils in dilute $\theta$ solution on the dynamics of the test polymer has been investigated. This was done within the Brinkman's phenomenology in which the solution is considered as a permeable medium, where the obstacles to the solvent flow are the polymer coils themselves. At sufficiently large concentrations of the coils the flow is effectively frozen and the polymer behaves as the Rouse chain even if the draining parameter is large. This hydrodynamic screening is not only concentration-dependent but the type of the polymer dynamics changes in the time as well.

In spite of the long-standing interest to the polymer physics, the creation of adequate theories of the polymer dynamics still represents a challenge. Even in the simplest case of flexible polymers in dilute solutions a number of problems remains unsolved and new questions appear. The presented work was inspired by the hydrodynamic theory of the Brownian motion and the computer experiments on simple liquids. Several decades ago these "experiments" markedly contributed to the understanding of the physics of liquids and initiated an enormous flow of investigations in this field. We believe that the peculiarities found here for the polymer motion could stimulate new studies towards a better understanding of the dynamics of polymers.

ACKNOWLEDGMENT

We are greatly indebted to Dr. O. Krichevsky for providing us with the experimental data [32]. This work was supported by the grant VEGA 1/0429/03, Scientific Grant Agency of the Slovak Republic.